\documentclass{aa}
\usepackage{natbib}
\usepackage{graphicx}
\usepackage{babel}
\usepackage{caption}
\usepackage{subcaption}
\usepackage[varg]{txfonts}
\usepackage{rotating}
\usepackage{dcolumn}
\usepackage{fmtcount}
\usepackage{longtable, booktabs}
\usepackage{array}
\usepackage{color}
\usepackage{aas_macros}
\bibpunct{(}{)}{;}{a}{}{,}
\usepackage{xcolor}
\definecolor{dark-red}{rgb}{0.5,0.15,0.15}
\definecolor{dark-blue}{rgb}{0.15,0.15,0.5}
\definecolor{medium-blue}{rgb}{0,0,0.5}
\definecolor{medium-red}{rgb}{1,0,0}
\hypersetup{
    colorlinks={true}, linkcolor={medium-red},
    citecolor={dark-blue}, urlcolor={medium-blue}
}
\newcommand{\kms}{km\,s$^{-1}$}
\renewcommand{\thesubfigure}{\arabic{subfigure}}
\newcommand{\teff}{$T_{\textrm{eff}}$}
\newcommand{\logg}{\textrm{log g}}
\newcommand{\feh}{\textrm{[Fe/H]}}

\begin{document}

\def\newblock{\hskip 0.10em plus 0.15em minus 0.05em}
\setlength{\bibsep}{0pt}

\title{New atmospheric parameters and spectral interpolator for the MILES cool stars}

\author{Kaushal Sharma
          \inst{1,2}
          \and
        Philippe Prugniel\inst{1}
          \and
        Harinder P. Singh\inst{2,1}
          }

\institute{Universit\'{e} de Lyon, Universit\'{e} Lyon 1, 69622 Villeurbanne;
              CRAL, Observatoire de Lyon, CNRS UMR 5574, 69561 Saint-Genis Laval, France\\
              \email{kaushal@physics.du.ac.in; philippe.prugniel@univ-lyon1.fr}
         \and
             Department of Physics and Astrophysics, University of Delhi,
             Delhi-110007, India\\
             \email{hpsingh@physics.du.ac.in}
             }

   \date{Received xxxx; accepted xxx}

\abstract
{
The full spectrum fitting of stellar spectra against a library of empirical spectra is a well-established approach to measure the atmospheric parameters of FGK stars with a high internal consistency. Extending it towards cooler stars still remains a challenge.
}
{
We address this question by improving the interpolator of the MILES (Medium-resolution INT Library of Empirical Spectra) library in the low effective temperature regime (\teff{}\,<\,4800\,K), and we refine the determination of the parameters of the cool MILES stars.
}
{
We use the ULySS package to determine the atmospheric parameters (\teff, \logg{} and \feh{}), and  measure the biases of the results
with respect to our updated compilation of parameters calibrated against theoretical spectra.
After correcting some systematic effects, we compute a new interpolator that
we finally use to redetermine the atmospheric parameters homogeneously and assess the biases.
}
   {
Based on an updated literature compilation, we determine 
\teff{}  in a more accurate and unbiased manner compared to those determined with the original interpolator. 
The validity range is extended downwards to about \teff{} $= 2900$\,K compared to $3500$\,K previously.
The mean residual biases on \teff{}, \logg{}, and \feh{}, with respect to the literature compilation  
for the coolest stars (\teff{}\,$\leq$\,3800\,K) computed using the new interpolator, are $-15$\,K, $-0.02$\,dex, and $0.02$\,dex respectively. The corresponding estimations of the external precision are 63\,K, 0.23 dex, and 0.15 dex respectively.
For the stars with \teff{} in the range 3800\,--\,4200\,K, the determinations of \teff{} and \feh{} have been slightly improved.
At higher temperatures, the new interpolator is comparable to the original one.
The new version of the interpolator is publicly available.}
   {}
   

\maketitle
   
\section{Introduction}\label{sec:Intro}
The libraries of stellar spectra, such as ELODIE \citep{PS1}, CFLIB \citep{valdes2004}, or MILES \citep{san2006}, are used for a variety of applications especially in the modelling of stellar populations \cite[e. g.][]{leborgne2004}. 
In that context, apart from the completeness and quality of these spectral databases \citep{singh2006}, the accurate calibration of stellar atmospheric parameters, temperature (\teff), surface gravity (\logg), and metallicity (\feh), is known to be critical \citep{prugniel2007,percival2009}. 
For instance, changing the temperature of the giant branch is similar to displacing the isochrones, and it has a strong effect on the age determination of stellar populations.

The currently available libraries generally contain a fair number of cool stars, but often their parameters are poorly determined.
Improving the situation is essential, for example, to constrain the initial mass function in star clusters or galaxies \citep{vanDokkum2012,conroy2012} from integrated spectra.

The classical methods used to derive the atmospheric parameters from high-resolution spectra consist in measuring the equivalent width of some well-chosen lines and comparing them to similar measurements of theoretical spectra. This represents the vast majority of the measurement in the 1996 edition of the compilation by \citet{cayrel1997}.
Besides this, fitting the spectra emerges as an alternative. Some examples of this approach are the TGMET code \citep{katz1998,soubiran2003}, MATISSE algorithm \citep{recioblanco2006}, MA$\chi$ \citep{jofre2010}, or iSpec \citep{blanco2014}.

Advantages of the spectrum fitting include its simplicity, which makes it easier to implement in automatic pipelines \citep[e.g.][]{worley2012}, and its robustness towards the noise and blending of the spectral lines. This technique has also been proven to be reliable at low resolution.
\citet{wu2011b} applied it at R~= $\lambda/\Delta\lambda \approx$ 5000, 
\citet[][hereafter PVK]{prugniel2011} at R~$\approx$ 2000.
\citet{koleva2012} have shown that the measurements remain reasonable even at R~$\approx$ 1000. 

Although spectrum fitting has mostly been used for FGK stars, its advantage is particularly clear for even cooler stars, where the lines are severely blended, and the continuum is hazardous to define. 
The goal of this paper is to check and improve the reliability of the determination of the parameters of stars cooler than \teff{}\,$=$\,4500\,K (K-M spectral types).

We follow the approach of PVK, who measured the atmospheric parameters of stars in the MILES library by comparing the spectra to the ELODIE spectral library. The MILES library has a spectral resolution R $\approx$ 2000 in the wavelength range 3536-7410 \AA, while the ELODIE library has R $\approx$ 42000 in the wavelength range 3900-6800 \AA{}. The parameter estimations were made using the full spectrum fitting, as implemented in ULySS \citep{koleva2009}. In this implementation, the minimization of the residuals between a target and a model spectrum provides estimates of the three stellar atmospheric parameters. The model spectrum is an interpolation over the reference library, and its quality relies on (i) the precision of the atmospheric parameters in the reference library, and (ii) the accuracy of the interpolator. For stars cooler than K5, the quality of determinations is poor because there are only a few cool stars in ELODIE and the parameters of these stars are not precisely determined. The comparison between the parameters measured in PVK and determinations compiled from the literature shows diverging biases at low temperature. Moreover, when the PVK interpolator is in turn used as a reference to study other spectra \citep[as in e. g. ][]{koleva2012}, the errors are propagated. In order to improve the characterization of cool stars and to enhance our capability to measure the parameters of cool stars, we build a new interpolator for MILES library. Rather than using the PVK parameters, we correct them for detected systematics and we supplement them with compiled values for the coolest stars. 

In Sect.~\ref{sec:literature}, we introduce our updated compilation of the atmospheric parameters from the literature for the cool stars in MILES. 
In Sect.~\ref{sec:assessment}, we use ULySS \citep{koleva2009} and the interpolator used in PVK to estimate the parameters and assess the biases. 
In Sect.~\ref{sec:revisedinterpolator}, we adopt the refined parameters and apply systematic corrections to produce a revised MILES interpolator that we validate in Sect.~\ref{sec:results}. 
Conclusions are presented in Sect.~\ref{sec:conclusions}.

\section{Literature compilation}\label{sec:literature}
   
In PVK, the atmospheric parameters were determined using the ELODIE library as a reference, 
and the external precision and biases were assessed by comparing them to the literature compilation 
of \citet{cenarro2007} and to the homogeneous series of measurements from \citet{elo31} 
and \citet{wu2011}. 
These atmospheric parameters were judged to be reliable over most of the parameter space, but restrictions apply to parameter regimes located at the margins of the regions populated by the library stars, namely the coolest stars and those with the lowest metallicity.
For these stars, we generally adopted atmospheric parameters compiled from the literature. 
As the goal of this paper is to improve the determination of the parameters and the quality of the MILES interpolator in the regime of the cool stars, the first task is to assemble an up-to-date compilation of their atmospheric parameters, that we subsequently use as a reference to measure the biases and precision of our own measurements.

We selected the 332 MILES stars with \teff{}\,$\leq$\,4800\,K in 
PVK or \citet{cenarro2007} and searched the literature, and in particular the Pastel database \citep{pastel2010}, for recent analyses of their atmospheric parameters. 
Although the focus of the paper is on stars cooler than  \teff  $\lesssim 4500$\,K, we set the limit to a 
somewhat warmer value, to establish continuity with the whole sample.
We excluded a carbon star, HD\,187216 (MILES~720), which is not relevant in the present context.
Therefore, our sample contains 331 stars.

It is known that different series of measurements differ by systematic biases, in particular, due to the adoption of different sets of reference theoretical spectra (involving different physical ingredients) or to the usage of different spectral features. Unfortunately, it is not possible to perform an ad-hoc homogenization, as for example in \citet{cenarro2007}, for the warmer stars, where large series of measurements are inter-compared and corrected for systematics. This is because the measurements for cool stars are still scarce in the literature and are often available for only one or a few MILES stars in a given article. Still, we found a significant number of new measurements, which were not available at the time of the previous compilations. 
This enabled us to carry out a critical analysis of those data and to adopt our best estimate of the parameters. The adopted parameters are listed in Table~\ref{tab:params}. 
The sixth column in this table depicts a compilation quality
flag, which is labelled "0" when more than six reliable \feh{} measurements are found, "1" when there are at least two measurements consistent within 0.30 dex, and "2" when there is a single spectroscopic measurement, or the measurements are not consistent, or the estimate is derived from a photometric calibration. 
The classification is based on \feh{} because accurate \teff{} and \logg{} are a prerequisite for those measurements.
In total 56 stars have a quality flag of "0", 136 have "1", and 139 have "2". 
For ten stars, we could not find any value of \feh{} in the literature. Eight of these are cool giants (\teff{}\,$\leq$\,3800\,K), one is dwarf (\teff{} $\sim$ 4100\,K), and the last one is a warmer giant.
In the updated compilation, the temperature of 15 stars slightly exceeds the initial limit of 4800~K.

\onltab{   
   \onecolumn
   \begin{longtab}\tiny
     \begin{center}
       \setlength{\tabcolsep}{7.0pt}
\begin{longtable}{@{}lcllrclrrcc@{}}
         \caption{
Compiled and measured atmospheric parameters for the 331 MILES cool stars.}\label{tab:params}\\
         \hline\hline

   Name &  Miles & \multicolumn{4}{c}{Literature compilation} & \multicolumn{4}{c}{Measurements}\\
        &        & \teff{} & \logg{} & \feh{} & C & \multicolumn{1}{c}{\teff{} $\pm$ error} & \multicolumn{1}{c}{\logg{} $\pm$ error} & \multicolumn{1}{c}{\feh{} $\pm$ error} & W & S \\

      &     & \multicolumn{1}{c}{(K)} & \multicolumn{1}{c}{(dex)} & \multicolumn{1}{c}{(dex)} & flag & \multicolumn{1}{c}{(K)} & \multicolumn{1}{c}{(dex)} & \multicolumn{1}{c}{(dex)} & flag & flag\\
       \hline
       \endfirsthead
       \caption{continued.}\\
       \hline\hline
   Name &  Miles & \teff{} & \logg{} & \feh{} & C & \teff{}$\pm$error & \logg{} $\pm$ error & \feh{} $\pm$ error & W & S\\

     &     & \multicolumn{1}{c}{(K)}  & \multicolumn{1}{c}{(dex)} & \multicolumn{1}{c}{(dex)} & flag & \multicolumn{1}{c}{(K)} & \multicolumn{1}{c}{(dex)} & \multicolumn{1}{c}{(dex)} & flag & flag \\       
\hline
        \endhead
       \hline
       \endfoot
       HD225212                       & 002  & 4152  &  0.90  & $  0.08$ &  1 & 	4117 $\pm$   49  &  0.68 $\pm$    0.15  &$  0.14 \pm  0.07~~$ &  0 &  0 \\
HD249                          & 005  & 4850  &  2.96  & $ -0.15$ &  2 & 	4731 $\pm$   93  &  2.83 $\pm$    0.22  &$ -0.31 \pm  0.09~~$ &  0 &  0 \\
HD448                          & 009  & 4840  &  2.83  & $  0.09$ &  1 & 	4770 $\pm$   98  &  2.61 $\pm$    0.24  &$  0.02 \pm  0.09~~$ &  1 &  0 \\
HD1326B                        & 012  & 3505  &  5.08  & $ -0.50$ &  1 & 	3563 $\pm$   35  &  4.85 $\pm$    0.07  &$ -0.57 \pm  0.16~~$ &  0 &  1 \\
HD3008                         & 019  & 4204  &  0.44  & $ -1.92$ &  0 & 	4364 $\pm$   77  &  0.68 $\pm$    0.14  &$ -1.83 \pm  0.11~~$ &  2 &  2 \\
HD3574                         & 024  & 3915  &  1.44  & $ -0.01$ &  2 & 	4019 $\pm$   23  &  1.13 $\pm$    0.12  &$  0.01 \pm  0.04~~$ &  0 &  0 \\
HD4656                         & 032  & 3912  &  1.45  & $ -0.14$ &  1 & 	3934 $\pm$   24  &  1.67 $\pm$    0.14  &$ -0.13 \pm  0.06~~$ &  0 &  1 \\
HD4744                         & 033  & 4569  &  2.20  & $ -0.82$ &  2 & 	4590 $\pm$   93  &  2.32 $\pm$    0.26  &$ -0.74 \pm  0.11~~$ &  0 &  0 \\
\hline
     \end{longtable}
     \end{center}
\tablefoot{
Columns 3-6 contain the parameters compiled from the
literature. C is the compilation quality flag defined in Sect.~\ref{sec:literature} (from 0 for the best data, to 2 for the poorest). 
Columns 7-9 present the determined parameters and their associated external errors and flags. W is the wavelength consistency flag defined in  Sect.~\ref{sec:atm_params} (0 are the solutions most consistent over the whole wavelength range).
S is the stability flag defined in  Sect.~\ref{sec:validation_assess} (0 is the most stable).
Unreliable metallicities are denoted with a colon ( refer to Sect.~\ref{sec:mgiants} and Sect.~\ref{sec:outliers} for more details)
\tablefoottext{a}{Altered designation or special notes reported in Table~\ref{tab:remarks}.}\\
} 
 \end{longtab} 

\twocolumn
}

\subsection{Metallicity of the star clusters}
\label{sec:compil-clusters}

\begin{table}
\caption{Compilation of the metallicity of Galactic globular clusters}
\label{tab:compil_globularclusters}
\scalebox{0.94}
{
\begin{tabular}{lrll}
\hline\hline
\multicolumn{1}{c}{Cluster} & \multicolumn{1}{c}{\feh{}} & \multicolumn{1}{c}{Reference}& \multicolumn{1}{c}{N} \\
\hline
NGC~288  &$-1.07$&\citet{cenarro2007}&c\\ 
NGC~288  &$-1.32\pm0.02$&\citet{carretta2009}&c\\
NGC~288  &$-1.22$&\citet{carretta-paper7}&110\\
NGC~288  &$-1.31$&\citet{carr2009}&10\\
{\it NGC~288  }&${\it -1.32}$&{\it \citet{harris2010}}&{\it c}\\
\hline
NGC~1904 &$-1.37$&\citet{cenarro2007}&c\\
NGC~1904 &$-1.58\pm0.02$&\citet{carretta2009}&c\\
NGC~1904 &$-1.54$&\citet{carretta-paper7}&58\\
NGC~1904 &$-1.58$&\citet{carr2009}&10\\
{\it NGC~1904 }&${\it -1.60}$&{\it \citet{harris2010}}&{\it c}\\
\hline
NGC~5272 &$-1.34$&\citet{cenarro2007}&c\\
NGC~5272 &$-1.48\pm0.03$&\citet{sakari2013}&ILS\\
NGC~5272 &$ -1.32$&\citet{harris2010}&c\\
{\it NGC~5272 }&${\it-1.50\pm0.05}$&{\it \citet{carretta2009}}&{\it c}\\
\hline
NGC~5904 &$-1.11$&\citet{cenarro2007}&c\\
NGC~5904 &$-1.33\pm0.02$&\citet{carretta2009}&c\\
NGC~5904 &$-1.35$&\citet{carretta-paper7}&136\\
NGC~5904 &$-1.34$&\citet{carr2009}&14\\
NGC~5904 &$-1.27$&\citet{gratton2013}&30\\
{\it NGC~5904 }&${\it -1.29}$&{\it \citet{harris2010}}&{\it c}\\
\hline
NGC~6121 &$-1.19$&\citet{cenarro2007}&c\\ 
NGC~6121 &$-1.18\pm0.02$&\citet{carretta2009}&c\\ 
NGC~6121 &$-1.20$&\citet{carretta-paper7}&103\\ 
NGC~6121 &$-1.17$&\citet{carr2009}&14\\ 
NGC~6121 &$-1.07$&\citet{malavolta2014}&322\\ 
NGC~6121 &$-1.16$&\citet{malavolta2014}&1869\\ 
{\it NGC~6121 }&${\it -1.16}$&{\it \citet{harris2010}}&{\it c}\\ 
\hline
NGC~6205 &$-1.39$&\citet{cenarro2007}&c\\
NGC~6205 &$-1.58\pm0.04$&\citet{carretta2009}&c\\
NGC~6205 &$-1.53\pm0.02$&\citet{sakari2013}&ILS\\
{\it NGC~6205 }&${\it -1.53}$&{\it \citet{harris2010}}&{\it c}\\
\hline
NGC~6341 &$-2.16$&\citet{cenarro2007}&c\\
NGC~6341 &$-2.35\pm0.05$&\citet{carretta2009}&c\\
{\it NGC~6341 }&${\it -2.31}$&{\it \citet{harris2010}}&{\it c}\\
\hline
NGC~6838 &$-0.84$&\citet{cenarro2007}&c\\
NGC~6838 &$-0.82\pm0.02$&\citet{carretta2009}&c\\
NGC~6838 &$-0.80$&\citet{melendez2009}&9\\
NGC~6838 &$-0.81$&\citet{carretta-paper7}&39\\
NGC~6838 &$-0.83$&\citet{carr2009}&12\\
{\it NGC~6838 }&${\it -0.78}$&{\it \citet{harris2010}}&{\it c}\\
\hline
\end{tabular}
}
\tablefoot{
The adopted value is the last of the series for each cluster, emphasized in italic.
The last column, labelled N gives the number of stars observed in the cluster, ILS for integrated-light spectroscopy, or c for the compilations.
}
\end{table}

Fifty of our 331 stars are presumably members of star clusters.
For these, we adopt the metallicity of the cluster, established from detailed spectroscopic analysis of a number of stars averaged together, rather than individual measurements.
We initially used the compilation from \citet{carretta1997}, such as in PVK, but decided to switch to the metallicity scale of \citet{carretta2009}. This new scale appeared more consistent with the metallicity of the field stars.

After searching the literature, we adopted the metallicities compiled by \citet{harris2010}, which is an updated version of \citet{harris}. While the original catalogue was set on the \citet{ZW1984} metallicity scale (based on photometric and spectro-photometric indices), the new version adopted the \citet{carretta2009} scale. 
In the case of NGC~5272 (M~3), however, the \citet{harris2010} metallicity appears to significantly differ from \citet{carretta2009} and other recent measurements. For this cluster, we adopted \citet{carretta2009}.

Table~\ref{tab:compil_globularclusters} compiles the values of \feh{} for the Galactic globular clusters observed in MILES.
In Table~\ref{tab:compil_openclusters} we list \feh{} measurements of open clusters, in particular, those taken from the compilation by \citet{pancino2010} and \citet{heiter2014}.
The latter gathers high-resolution spectroscopic measurements of member stars and produces an average \feh{} for each cluster. We adopt these values for all but two of the clusters. For IC~4725, which lacks any high-resolution spectroscopy, we adopt the photometric metallicity obtained by \citet{netopil2013}. 
In both tables, the value adopted for each cluster is on the last line.

NGC~2420 was considered as one of the metal-poorest open cluster until it was revised to a near-solar value \citep[see][]{heiter2014}. This cluster is one of those used for the calibration of the SDSS \citep{lee2008,smolenski2011}, and to clarify the status of this cluster, we analysed the SDSS spectra of 90 presumed member stars, using ULySS, and we found \feh{}\,$=\,-0.34$\,dex.
This casts serious doubts about the revision towards a solar metallicity of the cluster and therefore we adopt the value from \citet{smolenski2011}. 

For two of the five open clusters, NGC~2682 = M~67 and NGC~6791, our adopted metallicities agree with \citet{cenarro2007}, while for the others, the revision is sensible.

\begin{table}
\caption{Compilation of the metallicity of Galactic open clusters}
\label{tab:compil_openclusters}
\scalebox{0.95}
{
\begin{tabular}{lrll}
\hline\hline
\multicolumn{1}{c}{Cluster} & \multicolumn{1}{c}{\feh{}} & \multicolumn{1}{c}{Reference}& \multicolumn{1}{c}{N} \\
\hline
IC~4725  &$ 0.17$&\citet{cenarro2007}&c\\
{\it IC~4725  }&${\it  0.03\pm0.08}$&{\it \citet{netopil2013}}&{\it p}\\
\hline
NGC~2420 &$-0.44$&\citet{cenarro2007}&c\\
NGC~2420 &$-0.04\pm0.03$&\citet{pancino2010}&3\\
NGC~2420 &$-0.20\pm0.06$&\citet{jacobson2011}&9\\
NGC~2420 &$-0.05\pm0.02$&\citet{heiter2014}&c\\
{\it NGC~2420 }&${\it -0.31}$&{\it \citet{smolenski2011}}&{\it 164}\\
\hline
NGC~2682 &$ 0.02$&\citet{cenarro2007}&c\\
NGC~2682 &$ 0.05\pm0.02$&\citet{pancino2010}&3\\
NGC~2682 &$-0.03\pm0.02$&\citet{tautvaisiene2000}&10\\
NGC~2682 &$ 0.02\pm0.04$&\citet{yong2005}&3\\
NGC~2682 &$ 0.03\pm0.02$&\citet{randich2006}&10\\
NGC~2682 &$ 0.03\pm0.04$&\citet{pace2008}&6\\
{\it NGC~2682 }&${\it  0.00\pm0.06}$&{\it \citet{heiter2014}}&{\it c}\\
\hline
NGC~6791 &$ 0.40$&\citet{cenarro2007}&c\\
NGC~6791 &$ 0.32\pm0.02$&\citet{worthey2003}&16\\
NGC~6791 &$ 0.46\pm0.08$&\citet{gratton2006}&3\\
NGC~6791 &$ 0.41\pm0.03$&\citet{geisler2012}&5\\
NGC~6791 &$ 0.30\pm0.02$&\citet{boesgaard2015}&40\\
{\it NGC~6791 }&${\it  0.42\pm0.05}$&{\it \citet{heiter2014}}&{\it c}\\
\hline
NGC~7789 &$-0.13$&\citet{cenarro2007}&c\\
NGC~7789 &$ 0.04\pm0.07$&\citet{pancino2010}&3\\
NGC~7789 &$-0.03\pm0.05$&\citet{tautvaisiene2005}&7\\
{\it NGC~7789 }&${\it  0.01\pm0.04}$&{\it \citet{heiter2014}}&{\it c}\\
\hline
\end{tabular}
}
\tablefoot{
The adopted value is the last of the series for each cluster.
The last column, labelled N gives the number of stars observed in the cluster, or c for the compilations and p for the photometric determinations.
}
\end{table}

\section{Assessment of the biases}\label{sec:assessment}

In this section, we first compare the parameters derived in PVK with our new compilation and discuss the biases.
As a second step, we use the interpolator ${\rm TGM}\,(\,T_{\rm{eff}},\log g,{\rm [Fe/H]},\lambda)$ based on the MILES spectra and the PVK parameters to redetermine the atmospheric parameters and compare them with our compilation.
This step accumulates the effects of the biases in the PVK parameters to additional possible systematics introduced by the interpolator.
The interpolator presented in PVK is referred to as V1, and the improved version that we build in Sect.~\ref{sec:revisedinterpolator} is called V2.

\subsection{Biases in PVK}\label{sec:biases_pvk}

Figure~\ref{fig:biases} (left column) compares the parameters measured with the ELODIE interpolator and published in PVK with those of our compilation(some stars from the current sample were not measured in PVK, because they are out of the validity range of the ELODIE interpolator). The red lines in the lower panels, which show the residuals (PVK $-$ compilation vs. our compilation), are running averages revealing the biases. 
The bold sections of these lines indicate the regions where the bias is considered significant, i.e. where locally the bias exceeds three times its standard deviation. 
Although they remain within reasonable limits, significant biases are detected for all three parameters. They are of about 40~K on \teff{}, 0.20~dex on \logg{} and 0.10~dex on \feh.

The \teff{} determinations are unbiased around 3800 K. For the cooler giants, however, the temperature becomes increasingly underestimated, reaching a bias of about 150\,K at 3300\,K. At variance, the temperature of the dwarfs is overestimated by about 100\,K at the same point. The net effect seen on Fig.~\ref{fig:biases} is dominated by the giants, which are more numerous. 
At warmer temperature ($4000 \leq$~\teff{}~$\leq 4400$\,K), one notes an opposite positive bias of about 40~K, with a similar contrast between the dwarf and the giants.
The divergence at low temperature certainly reflects the lack of very cool stars in the ELODIE library, and in fact PVK did not provide the measurements for the coolest stars (missing on Fig.~\ref{fig:biases}).

The main significant patterns in the comparison of \logg{} between the PVK measurements and the literature are the concentrations of measurements near \logg{} $\approx$ 2.6 and 4.6 dex, corresponding to compiled values in the range 2 to 3 dex, and 4 to 5 dex, respectively (another, less marked, concentration occurs around \logg{} $\approx$ 1.6 dex). Expressed differently, for these groups of stars, the PVK measurements are less dispersed than the literature values. Given the limited precision of the literature measurements and their lack of homogeneity, it is presently difficult to interpret this effect.

The metallicities measured by PVK generally appear larger by about 0.1~dex with respect to the literature, except for the cool dwarfs whose metallicity is dramatically underestimated.

Statistics of the comparisons between PVK and the compilation is presented in Table~\ref{tab:biases}. 
The table gives different regions of the parameter space where systematics were investigated. The statistics were computed with a robust estimator (see the IDL procedure \textsc{biweight\_mean}\footnote{Part of the IDL Astronomy User's Library available at \\
\url{http://idlastro.gsfc.nasa.gov}}) to minimize the effect of the outliers.

\subsection{Biases in remeasured parameters}
\label{sec:measV1}

In PVK, the estimated parameters of MILES stars were used to build the V1 interpolator for the MILES library. The principle of an interpolator is to represent each wavelength bin of the library with a function of \teff, \logg{} and \feh. This allows one to compute a so-called interpolated spectrum for any \teff, \logg, and \feh. 
The interpolated spectra can be used to compute stellar population models, as in \citet{leborgne2004}, or to measure the atmospheric parameters of an individual star from its spectrum, as in \citet{wu2011b}.
The V1 interpolator has the same form as that used for ELODIE 3.2 in \citet{wu2011b}.
It is split in three temperature regimes (hot, warm and cool stars) and each \teff{} range consists of a polynomial development of 23 terms combining the three parameters raised to various exponents. The coefficients of these polynomials are fitted over the whole library.

We use ULySS to determine the atmospheric parameters from the MILES (version 9.1, discussed in Appendix \ref{appendix:miles91}) spectra. This flexible
programme allows one to fit virtually any kind of non-linear model (or constrained linear combinations of 
these models) to a spectrum. In the present case, we use the TGM component provided with
the package, feeding it with the MILES interpolator. The model fitted to each MILES spectrum $S(\lambda)$ can be written as \citep{wu2011b}
   \begin{equation}
     {S(\lambda)} = P_{n}(\lambda) \times [\,{\rm TGM}\,(\,T_{\rm{eff}},\log g,{\rm [Fe/H]},\lambda) \otimes G(\,v_{sys},\sigma)],
   \end{equation}
where 
$P_{n}(\lambda)$ is a series of Legendre polynomials up to the degree $n$, and $G(v_{sys},\sigma)$ 
is a Gaussian broadening function characterized
by the systemic velocity $v_{sys}$, and the dispersion $\sigma$. 
ULySS minimizes the squared residuals between the observed spectrum
and  $S(\lambda)$.
The free parameters of the minimization procedure are those of the 
TGM function, \teff, \logg, and \feh;
the two parameters of the broadening function, $v_{sys}$ and $\sigma$; and the
coefficients of $P_{n}$. The parameter $v_{sys}$ absorbs the imprecision of the
catalogued radial velocity of the stars that were used to reduce them
in the rest frame; $\sigma$ encompasses both the dispersion of instrumental
broadening and the effect of rotation. The Legendre polynomials
absorb the uncertainties in the flux calibration, which is normally
excellent in MILES, and on the corrections of the Galactic extinction, which is 
not always accurately known.   
This procedure was previously used by PVK, \citet{wu2011, wu2011b},
and \citet{koleva2012}.

We tried to fit our sample in different spectral ranges, including the whole range or restricting it in the blue or the red. 
We found that excluding the blue range, below 4200 \AA{}, reduces the mean dispersion 
between the solution and compiled parameters. Using the blue end quadratically adds an error of about 45~K on \teff{}, increasing the mean external error from 70 to 83~K (for the whole sample). 
We did not fully investigate the reasons for this effect, but we can {\it a priori} exclude that it is due to the lower signal-to-noise ratio (S/N) in the blue. 
Indeed, the S/N remains generally larger than 30, and, as shown in Sect.~\ref{sec:atm_params}, this would affect the errors by a significantly smaller amount. 
A more likely cause is the high sensitivity of the blue region to the diversity of abundances of the various chemical elements \citep{marcum2001,koleva2012} in the stars of the library. 
Notwithstanding a robust explanation, we restricted the fitting range to 4200 \AA{} in blue (PVK adopted the same limit).
The red part of the spectra above 5880~\AA{} (i.e. the NaD doublet), are plagued with strong telluric absorption lines due to $\rm{H_2O}$ or $\rm{O_2}$, which were corrected in MILES. However, these corrections have necessarily a limited precision, resulting in misfits for some spectra. However, clipping the entire red region does not improve the consistency between our solution and the compilation and, therefore, we kept it, masking only the most affected NaD feature.
Since we measure the iron metallicity, 
we also masked the Mg$_b$ feature (5167-5184\,\AA), which is the most prominent signature of $\alpha$-elements abundance. However, we did not detect significant effect on the solution.
Both the blue region and the Mg$_b$ feature would be naturally useful if we wanted to measure both [Mg/Fe] and \feh, as in \citet{prugniel2012}.

\begin{figure*}
\begin{center}
\includegraphics[width=0.82\textwidth]{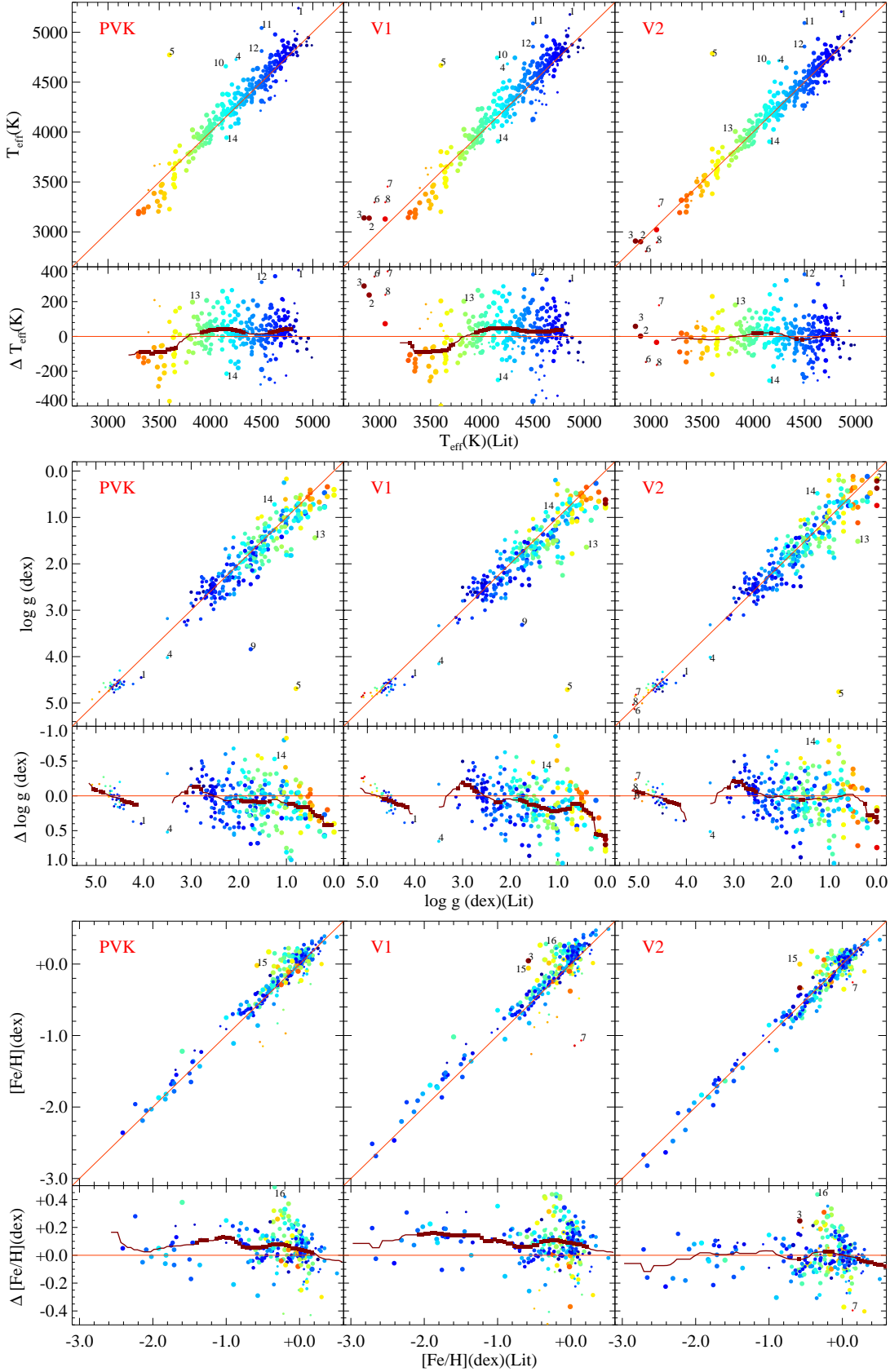}
\caption{Comparison of the atmospheric parameters adopted in PVK and estimated using
V1 and V2 versions of the interpolator with those of the literature compilation.
The abscissae are the values from the compilation, and the ordinates of the top 
plot of each panel are those from PVK, V1, or V2. The residuals presented in the bottom plot of each panel 
are the  PVK, V1, or V2 values minus those of the compilation. 
The red lines show the biases, the bold sections indicating regions where they are statistically significant. 
The stars labelled with numbers are discussed in Sect.~\ref{sec:outliers}. The colour of the symbols represents
the temperature and the size is linked to the surface gravity.
}
       \label{fig:biases}
     \end{center}
   \end{figure*}

In order to avoid possible local minima, we performed a global minimization with a grid of initial guesses, following \citet{wu2011b}.
We used the grid \teff{}~$\in [3000, 4500]$~K, \logg{}~$\in [1.0, 4.0]$~dex, and  \feh{}~$\in [-2.0, -0.5, 0.3]$~dex, and if the fit converged to different solutions, we selected the best one, corresponding to the global minimum. We found that starting from the compiled values as a single guess would have provided solutions within the error bars, but the solutions would not be formally independent from the compilation.
We used a maximum degree $n = 40$ for $P_{n}$, tested as described in \citet{wu2011b}, and rejected the spikes from the fit using ULySS' {\sc /clean} option.
For the cluster stars included in our selection, we fit only \teff{} and \logg{}, and adopt the metallicity compiled in Sect.~\ref{sec:compil-clusters} (the solution with the three parameters free is presented in Sect.~\ref{sec:clusters}). These cluster stars are not used in the statistics involving \feh{} and are not shown on the \feh{} plots of Fig.~\ref{fig:biases}.

As shown in Fig.~\ref{fig:biases} (central column labelled V1), the features are qualitatively similar to those of the left column, which presents the comparisons with the PVK parameters. At the lowest temperature, the biases are the same, negative for the giants, and positive for the dwarfs. For warmer stars, the bias increased by about 50\% with respect to PVK.
This degradation suggests that the biases may partly be due to the analytical form of the interpolator. While the measurements in PVK suffer from the interpolator's bias, the remeasured parameters are affected twice by the effect. The bias on \feh{} suffered a similar amplification, while those on \logg{} are basically unchanged.

\begin{table*}[!ht]
     \begin{center}
       \caption{Comparison statistics of the derived atmospheric parameters with literature compilation.}\label{tab:biases}
       \begin{tabular}{lccrrrrrrr}\hline\hline
         \\
         Region     & $N$  &     &\multicolumn{2}{c}{$\Delta$\,\teff{}~(K)}&
         \multicolumn{2}{c}{$\Delta$\,\logg{} (dex)} &\multicolumn{3}{c}{$\Delta$\,\feh{}(dex)}\\
                               &     &      &\multicolumn{1}{c}{$\mu$} & \multicolumn{1}{c}{$\sigma$} & 
         \multicolumn{1}{c}{$\mu$} & \multicolumn{1}{c}{$\sigma$}  & 
         \multicolumn{1}{c}{$N$} & \multicolumn{1}{c}{$\mu$} & \multicolumn{1}{c}{$\sigma$} \\\hline
                                                                                 & 33 & PVK &  -94 &  115 &   0.07 &   0.31 &   27 &   0.07 &   0.22\\
                                \teff{}\,$\leq$\,3800\,K and \logg{}\,<\,3.5 dex & 40 &  V1 &  -81 &  134 &   0.18 &   0.37 &   30 &   0.12 &   0.27\\
                                                                                 & 40 &  V2 &  -15 &   82 &   0.00 &   0.38 &   30 &   0.04 &   0.23\\
\hline
                                                                                 & 77 & PVK &   42 &   81 &   0.12 &   0.34 &   61 &   0.11 &   0.12\\
                       3800\,<\,\teff{}\,$\leq$\,4200\,K and \logg{}\,<\,3.5 dex & 77 &  V1 &   43 &   90 &   0.11 &   0.32 &   61 &   0.16 &   0.13\\
                                                                                 & 77 &  V2 &   17 &   74 &   0.03 &   0.31 &   61 &   0.02 &   0.14\\
\hline
                                                                                 & 69 & PVK &   21 &   88 &   0.09 &   0.26 &   53 &   0.02 &   0.09\\
                       4200\,<\,\teff{}\,$\leq$\,4500\,K and \logg{}\,<\,3.5 dex & 72 &  V1 &   49 &  113 &   0.15 &   0.28 &   55 &   0.08 &   0.10\\
                                                                                 & 72 &  V2 &   -1 &   80 &   0.07 &   0.25 &   55 &  -0.04 &   0.10\\
\hline
                                                                                 & 86 & PVK &   42 &   69 &   0.04 &   0.25 &   72 &   0.05 &   0.08\\
                       4500\,<\,\teff{}\,$\leq$\,4800\,K and \logg{}\,<\,3.5 dex & 89 &  V1 &   45 &   77 &   0.06 &   0.25 &   75 &   0.08 &   0.09\\
                                                                                 & 89 &  V2 &    4 &   77 &  -0.01 &   0.25 &   75 &   0.01 &   0.09\\
\hline
                                                                                 &  3 & PVK &  137 &  105 &  -0.11 &   0.11 &    3 &  -0.66 &   0.12\\
                           \teff{}\,$\leq$\,3800\,K and \logg{}\,$\geq$\,3.5 dex &  9 &  V1 &  166 &  132 &  -0.14 &   0.12 &    9 &  -0.72 &   0.39\\
                                                                                 &  9 &  V2 &  -13 &  120 &  -0.06 &   0.12 &    9 &  -0.04 &   0.16\\
\hline
                                                                                 & 28 & PVK &  -45 &  120 &   0.02 &   0.16 &   27 &  -0.08 &   0.19\\
                  3800\,<\,\teff{}\,$\leq$\,4800\,K and \logg{}\,$\geq$\,3.5 dex & 29 &  V1 & -112 &  140 &   0.07 &   0.14 &   28 &  -0.03 &   0.15\\
                                                                                 & 29 &  V2 &  -35 &  123 &   0.02 &   0.14 &   28 &  -0.03 &   0.14\\
\hline

         \hline
\end{tabular}
\end{center}
\tablefoot{For each parameter $\mu$ and $\sigma$
are the mean and dispersion of the differences in the derived parameters.
For each range of \teff{}, the first row
is the mean and sigma of biases in the parameters in PVK with respect to literature compilation. 
The other two rows are the mean and dispersion in the derived parameters using two versions of the 
interpolator. The robust statistics has been computed with the IDL command \textsc{biweight\_mean}.
Cluster stars were not included in the \feh{} statistics.
}
\end{table*}

\section{Revision of the interpolator}\label{sec:revisedinterpolator}

In PVK, the parameters of the MILES stars were derived using the interpolator described in \citep{wu2011b}, and were used to build the V1 interpolator. In the previous section, we used V1 to derive a new set of parameters, and the analysis of the two sets revealed some limitations and residual biases. In the present section, we build a new interpolator for MILES, V2, with improved input parameters (correcting biases) and extended validity range.

\subsection{Input catalogue}\label{sec:input_cat}

The precision and consistency of the input catalogue is crucial to the quality
of the interpolator. The biases seen in Fig.~\ref{fig:biases} for the PVK parameters unavoidably propagate to
the interpolator computed with those parameters, and then to subsequent measurements of the
atmospheric parameters.

The use of the compiled parameters should avoid the biases detected above, but those parameters have lower internal consistency than PVK.
This may introduce other artifacts in the interpolation. We therefore adopt a hybrid solution:
\begin{itemize}
\item[-]In the regions at the border of the parameter space, the PVK values are not reliable and we adopt the compiled values. This concerns the coolest stars, and those with the lowest metallicity.
\item[-]In the well-populated regions of the parameter space, we identified
systematic effects in the PVK values and we corrected them. Rather than using
the compiled values, this procedure preserves the internal consistency of PVK.
\end{itemize}

\subsection{Extension towards cool dwarfs}
\label{sec:extensioncooldwarfs}

The computation of the interpolator involves some extrapolation support spectra
intended to extend the validity range on the margin of the populated region of the parameter space (see details in PVK).

To improve the interpolator in the M stars regime, we extracted high-resolution spectra from the FEROS archive for 24 of the 102 stars whose atmospheric parameters were homogeneously determined by \citet{neves2013}.
Their spectral types lie in the range M0 to M4.5.
FEROS is an echelle spectrograph attached to the 2.2 m ESO telescope and its 
wavelength coverage encompasses the range of MILES. The archived spectra
were automatically processed for the standard CCD reduction up to the wavelength calibration and order connection stage. We reduced them to the velocity rest frame, clipped the spikes interactively, convolved to the resolution of MILES, and adjusted the flux calibration to be consistent with MILES.
Finally, we used these spectra in the interpolator, giving them a low weight (the FEROS spectra have all together the weight of three MILES spectra).
We used the metallicities determined by \citet{neves2013} and the temperatures computed from the V and K$_S$ colours with the \citet{casagrande2008} relation. The gravities were interpolated from the relation between the spectral type and parameters listed in \citet{allen1973}.
This allowed us to improve the quality of the interpolator for the cool dwarfs.

\begin{figure}[h]
 \centerline{\includegraphics[width=7.5cm,height=7.5cm]{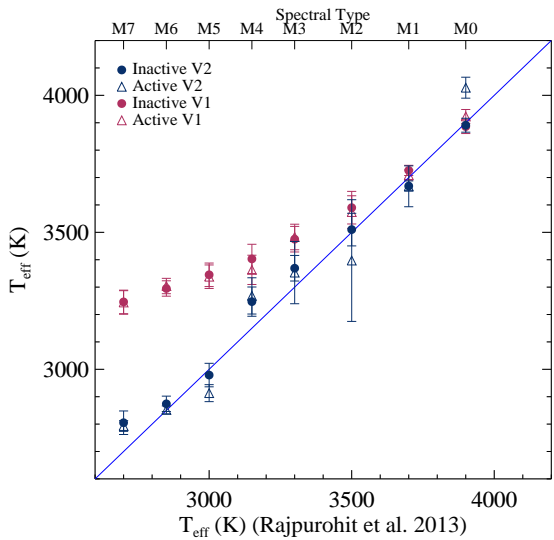}}
 \caption{
 Effective temperatures estimated with V1 and V2 interpolators for the \citet{hawley2007} M dwarf templates compared with the \teff{} scale of \citet{rajpurohit2013}. V1 and V2 solutions are shown in maroon and blue, respectively. 
Open triangles represent the chromospheric active templates and filled circles represent chromospheric inactive templates.
}
 \label{fig:m_templates}
\end{figure}

\begin{table*}[]
\begin{center}
\caption{Fit of M star templates}
\label{tab:mtemplates}
\begin{tabular}{l | c | ccr | ccr | ccr}\hline\hline
Spectral& & \multicolumn{3}{c}{Inactive\tablefootmark{a}} & \multicolumn{3}{c|}{Active\tablefootmark{a}} & \multicolumn{3}{c}{UDC} \\
type    & \teff\tablefootmark{b} & \teff & \logg & \feh & \teff & \logg & \feh & \teff & \logg & \feh \\
\hline
M0 &  3900 & 3891 & 4.62 & $ -0.27 $ & 4028 & 4.47 & $ -0.25 $ & -- & -- & \multicolumn{1}{c}{--}\\
M1 &  3700 & 3669 & 4.74 & $ -0.29 $ & 3668 & 4.81 & $ -0.32 $ & -- & -- & \multicolumn{1}{c}{--}\\
M2 &  3500 & 3510 & 4.87 & $ -0.44 $ & 3397 & 4.67 & $ -0.63 $ & 3449 & 4.89 & $-0.27$\\
M3 &  3300 & 3369 & 4.92 & $ -0.43 $ & 3353 & 4.90 & $ -0.80 $ & 3315 & 4.93 & $-0.34$\\
M4 &  3150 & 3247 & 4.91 & $ -0.33 $ & 3268 & 5.04 & $ -0.73 $ & 3157 & 4.95 & $-0.11$\\
M5 &  3000 & 2979 & 4.85 & $ -0.03 $ & 2913 & 4.87 & $ -0.03 $ & 2901 & 4.97 & $ 0.04$\\
M6 &  2850 & 2874 & 5.02 & $ -0.04 $ & 2852 & 4.97 & $ -0.04 $ & 2854 & 5.02 & $ 0.12$\\
M7 &  2700 & 2805 & 5.18 & $ -0.04 $ & 2793 & 5.10 & $  0.05 $ & 2850 & 5.00 & $ 0.34$\\
\hline
\end{tabular}
\end{center}
\tablefoot{
\tablefoottext{a}{\citet{hawley2007} gives two sets of spectral templates, one for the stars without detected chromospheric activity, and one for the active stars.}
\tablefoottext{b}{Spectral type vs. \teff{} relation from \citet{rajpurohit2013}, their figure 5.}
}
\end{table*}

We carried out a series of tests to validate the interpolator in this regime.
We inverted FEROS spectra for 45 dwarfs with spectral type between M0 to M4.5, including the 24 used above to constrain the interpolator.
The derived atmospheric parameters are unbiased with respect to the measurements given by \citet{neves2013}, and the standard deviations 
of the differences are 63 K on \teff, and 0.10 dex on \feh. These dispersions
can be compared to the 80 K and 0.08 dex errors estimated by \citet{neves2013} with their spectroscopic method on their calibration sample of 55 stars. The precision obtained here on \feh{} is possibly lower, but this can be attributed to the lower resolution and smaller number of calibrators.
The good correlation between the two series of measurements gives confidence in our metallicity measurements, down to \teff$\,=\,3000$\,K and in the range $-0.6 < \feh < 0.2$.

We also analysed the M dwarf templates from \citet{hawley2007}\footnote{downloaded from \url{http://www.astro.washington.edu/users/slh/}}. 
These spectra were built from the Sloan Digital Sky Survey (SDSS) by stacking thousands of individual observations. They have the approximate resolution of MILES. 
While the V1 solutions start to depart from the \citet{rajpurohit2013} scale after the spectral type M2 (see Table~\ref{tab:mtemplates} and Fig.~\ref{fig:m_templates}), the V2 solutions follow the relation up to M6. The validity range for the \teff{} determination has been extended from 3500\,K to 2900\,K. For the M7 template, the inverted temperature is biased upwards only by 100 K, which is a typical effect at the limit of the interpolator.
The residual spectra of the V1 fits indicated prominent misfits of the CaOH band (5500-5570 \AA), which have been corrected in V2.
Finally, we analysed 543 SDSS spectra of individual M0 to M7 stars taken from the 
Ultracool Dwarf catalogue \citep[UDC;][]{martin2005}. Together with the spectrum, the UDC gives the spectral classification taken from \citet{hawley2002}. After averaging the solutions of the fits for each spectral type, we find a good agreement
with the Rajpurohit's \teff{} scale for the M2 to M6 types (see Table~\ref{tab:mtemplates}). 
We did not report the results for the M0-M1 types because these classes count less spectra (the focus of the UDC is on very cool stars), and the fact that our determinations of the temperature fall about 150 K below the \citet{rajpurohit2013} scale for these subtypes is not significant. The UDC spectra have a low S/N, and unlike the former templates, which were stacked before the analysis, the solutions were averaged after the analysis. These results confirm the reliability of the \teff determinations down to about 2900~K, and illustrate the robustness of the method towards the noise.

\subsection{Tuning of the interpolator}

The correction of the input catalogue and the modifications at the margins of
the parameter space reduce the biases considerably when the interpolator is used to measure the parameters. 
However, some effects, such as a 40\,K bias for the stars with \teff{} in the range 4000-4300\,K, persisted. The fact that those patterns were already present in the ELODIE interpolator suggests that we should change the polynomial development. To add new terms we searched those contributing to the maximum reduction of the residuals between the observed and interpolated spectra. We found that using 26 terms in place of the 23 terms previously used significantly improved the modelling and decreased the biases.

We also changed the relative weight of the stars according to the temperature and width of the overlap region between the sets of polynomials for warm and cool stars. We fine-tuned these parameters to minimize the biases in the newly computed interpolator.

\subsection{Validity limits for M-type giant and supergiant stars}\label{sec:mgiants}

For the M-type dwarfs, Fig.~\ref{fig:m_templates} has shown that the \teff{} validity limit has been significantly extended downwards from V1 to V2. For the giants, Table~\ref{tab:biases} shows that the biases with respect to the literature have been reduced. As an additional test, we also check the spectral type vs. \teff{} relation in this section.

The temperatures of cool giants have been determined independent of spectroscopy, using
lunar occultations following \citet{ridgway1980,richichi1999} or interferometry \citep{dyck1996,perrin1998} to constrain the angular diameters. These fundamental calibrations were used to establish relations between the spectral type and temperature. We compare our \teff{} measurements with these calibrations.

To supplement MILES, a sequence of M giants with spectra available in the ELODIE archive \citep{moultaka2004}\footnote{\url{http://atlas.obs-hp.fr/elodie/}} is presented in Table~\ref{tab:refgiants}. For most of these stars, the ELODIE archive contains multiple observations; we analysed them individually and averaged the solutions.

\begin{figure}[h]
\centerline{\includegraphics[width=7.5cm]{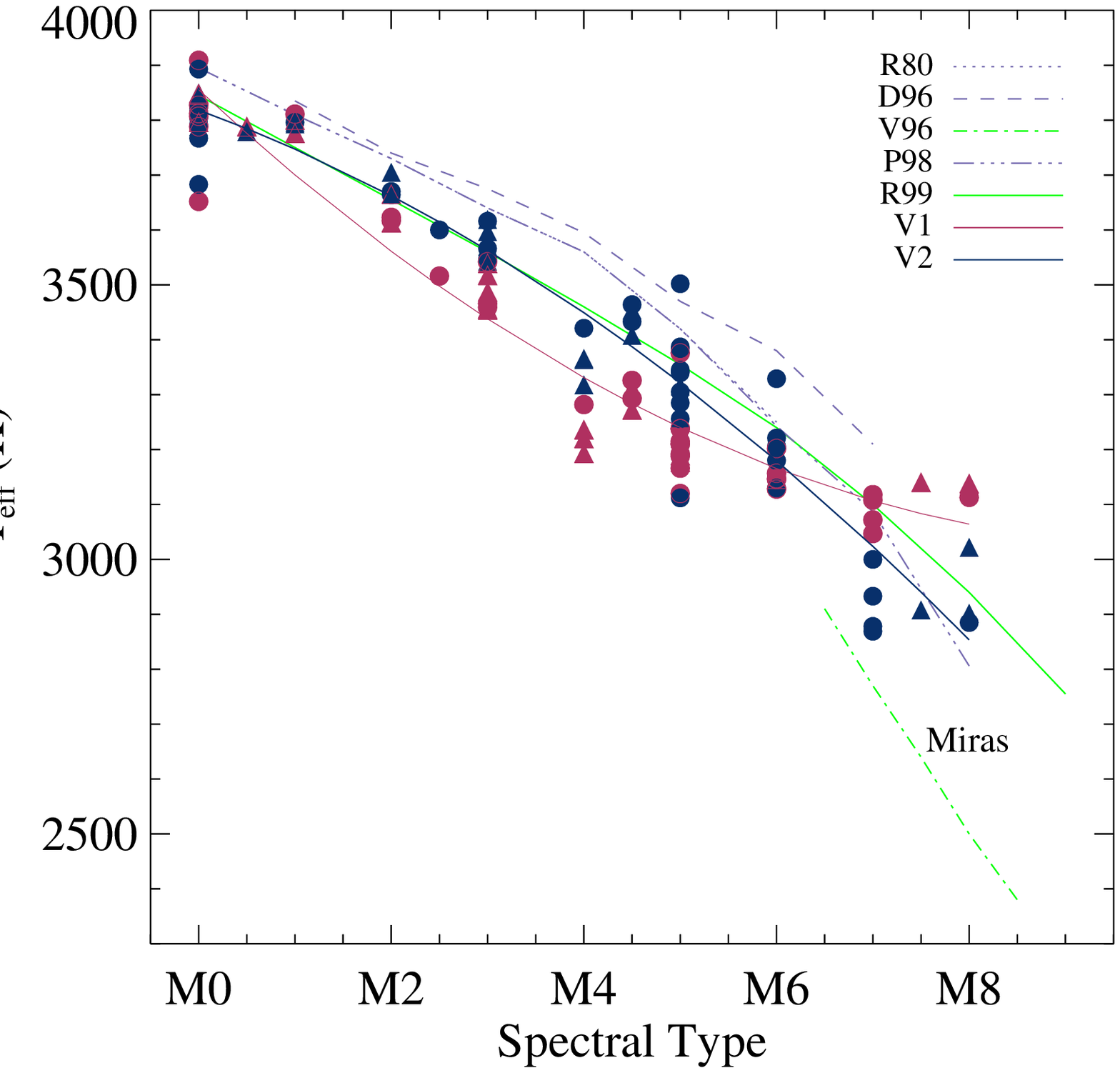}}
 \caption{ 
\teff{} vs. spectral type diagram for the M giants. The purple lines 
are the relations derived by
\citet[R80]{ridgway1980}, \citet[D96]{dyck1996}, \citet[P98]{perrin1998}.
The green lines are the relations derived by \citet[R99]{richichi1999} and \citet[V96]{vanbelle1996} as indicated in the legend.
The triangles represent the MILES spectra and the circles the ELODIE spectra.
V1 solutions are shown in maroon and V2 solutions in dark blue.
The maroon and dark blue lines are the quadratic fits to the V1 and V2 solutions, respectively. 
}
 \label{fig:m_giants}
\end{figure}

Figure~\ref{fig:m_giants} places the MILES and ELODIE spectra in the \teff{} vs. spectral type diagram. The quadratic fit shows that V1 solution is significantly more curved than the reference calibrations, restoring lower temperatures for the M3 to M5 types and then flattening to \teff $\approx$ 3100 K for the latest types. The V2 solution is close to the \citet{richichi1999} calibration. However, the latest M-type stars of the sequence are Miras, which are known to be cooler at the same spectral type \citep{vanbelle1996}. Our measurements do not restore these lower temperature.
We obtain positive bias of $\sim$\,100\,K for M7 and $\sim$\,400\,K for M8 type Miras with respect to the \citet{vanbelle1996} scale. We could not investigate this issue further because of the limited number of available spectra for these type of stars.
With V2 interpolator, we are able to extend the \teff{} validity limit for the coolest giants downwards to 2900\,K as compared to 3100\,K with V1. 

There are also nine M stars of luminosity class I or II in MILES (according to the classification reported in SIMBAD) and we found 16 additional stars with spectra available in the ELODIE archive. This sample spans the range of spectral types from M0 to M4.
We analysed the spectra for these 25 stars and found that the results match the relation between the spectral type and the temperature established by \citet{levesque2005}. 

In order to estimate the reliability of the \feh{} measurements for the giants, we searched the literature for detailed abundance measurements of these stars and retrieved their spectra from public telescope archives.
Out of the 36 stars for which we gathered spectra from ELODIE or MILES, only 11 are cooler than 3500\,K, and two cooler than 3400\,K (HD\,148783 and 163990). We did not find any correlation between our measured \feh{} and those from the literature for \teff$ < 3800$\,K.
This lack of correlation is in part due to the small range of metallicity span of these stars: the mean \feh{} is $-0.09$ dex in the literature and $-0.05$ dex in our measurements, with dispersions of $0.24$ and $0.19$ dex, respectively. This metallicity range is not larger than the expected precision of the measurements in the literature (as can be assessed from the dispersion between measurements of the same stars in different studies). 
Also, our MILES sample counts two cool giants from the NGC\,6791 cluster, MILES 940 and 941. While the former, at \teff~$\approx$ 3400\,K, is effectively inverted at nearly the cluster's metallicity (\feh\,= $-0.42$\,dex), the latter (\teff~$\approx$ 3200\,K) is measured at a near solar metallicity. This may indicate that the interpolators may be able to restitute super-solar metallicities at 3400\,K, but not at lower \teff.

Altogether, the very restricted number of cool giants with \feh{} available in our input catalogue and the limited precision of these measurements definitely affect the reliability of the interpolator in this regime.
We consider that the metallicity measurements of the giants cooler than 3800\,K are not fully reliable, and we flag these with a colon in Table~\ref{tab:params}.

\begin{table*}[]
\begin{center}
\caption{Sequence of cool giants from the ELODIE archive}
\label{tab:refgiants}
\begin{tabular}{l|l|l|l|c|c|r}\hline\hline
Name& HD/BD& SpClass\tablefootmark{a}  & Var. Type\tablefootmark{b} &\multicolumn{3}{c}{Measurements}\\
    &      &                           &                            & \multicolumn{1}{c}{\teff{} $\pm$ error} & \multicolumn{1}{c}{\logg{} $\pm$ error} & \multicolumn{1}{r}{\feh{} $\pm$ error\tablefootmark{c}} \\
\hline
$\mu$ Uma      &  HD089758      &  M0III         &  E:            & $ 3822 \pm 43$ & $  1.39 \pm 0.17$ & $ -0.20 \pm 0.06$ \\
$\alpha$ Vul   &  HD183439      &  M0III         &                & $ 3767 \pm 60$ & $  1.66 \pm 0.25$ & $ -0.37 \pm 0.12$ \\
gam Sge        &  HD189319      &  M0III         &                & $ 3893 \pm 55$ & $  1.56 \pm 0.17$ & $  0.03 \pm 0.06$ \\
$\beta$ And    &  HD006860      &  M0III         &  NSV?          & $ 3804 \pm 48$ & $  1.40 \pm 0.17$ & $ -0.26 \pm 0.06$ \\
HD 46784       &  HD046784      &  M0III         &                & $ 3683 \pm 50$ & $  1.19 \pm 0.25$ & $  0.02 \pm 0.12$ \\
106 Her        &  HD168720      &  M0III         &  SR:           & $ 3792 \pm 50$ & $  1.58 \pm 0.17$ & $ -0.08 \pm 0.06$ \\
55 Peg         &  HD218329      &  M1III         &  NSV?          & $ 3796 \pm 55$ & $  1.58 \pm 0.25$ & $  0.08 \pm 0.06$ \\
8 And          &  HD219734      &  M2III         &  NSV?          & $ 3670 \pm 40$ & $  1.01 \pm 0.25$ & $ -0.13 \pm 0.12$ \\
HD 42787       &  HD042787      &  M2III         &  SRB:          & $ 3664 \pm 52$ & $  1.14 \pm 0.25$ & $ -0.16 \pm 0.12$ \\
$\beta$ Peg    &  HD217906      &  M2.5II-IIIe   &  LB            & $ 3600 \pm 58$ & $  0.93 \pm 0.34$ & $ -0.35 \pm 0.18$ \\
104 Her        &  HD167006      &  M3III         &  SR            & $ 3616 \pm 55$ & $  0.99 \pm 0.34$ & $ -0.31 \pm 0.18$ \\
$\mu$ Gem      &  HD044478      &  M3III         &  LB            & $ 3566 \pm 43$ & $  0.68 \pm 0.25$ & $ -0.02 \pm 0.12$ \\
i Dra          &  HD121130      &  M3III         &  LB:           & $ 3547 \pm 50$ & $  0.80 \pm 0.25$ & $ -0.03 \pm 0.12$ \\
ome Vir        &  HD101153      &  M4III         &  LB            & $ 3421 \pm 35$ & $  0.48 \pm 0.25$ & $ -0.09 \pm 0.12$ \\
BY Boo         &  HD123657      &  M4.5III       &  LB:           & $ 3433 \pm 37$ & $  0.56 \pm 0.25$ & $ -0.11 \pm 0.18$ \\
V1743 Cyg      &  HD184786      &  M4.5III       &  SRB:          & $ 3464 \pm 30$ & $  0.44 \pm 0.17$ & $ -0.09 \pm 0.12$ \\
R Lyr          &  HD175865      &  M5III         &  SRB           & $ 3340 \pm 34$ & $  0.58 \pm 0.17$ & $ -0.19 \pm 0.12$ \\
Z Eri          &  HD017491      &  M5III         &  SRB           & $ 3285 \pm 42$ & $  0.72 \pm 0.17$ & $ -0.16 \pm 0.12$ \\
AR Cet         &  HD012292      &  M5III         &  SR:           & $ 3342 \pm 40$ & $  0.63 \pm 0.17$ & $ -0.10 \pm 0.12$ \\
SS Cep         &  HD022689      &  M5III         &  SRB           & $ 3112 \pm 55$ & $  0.82 \pm 0.25$ & $ -0.27 \pm 0.12$ \\
FL Cam         &  HD074225      &  M5III         &  LB            & $ 3384 \pm 37$ & $  0.44 \pm 0.25$ & $ -0.07 \pm 0.12$ \\
CZ Lyn         &  HD076386      &  M5III         &  SRB           & $ 3502 \pm 45$ & $  0.56 \pm 0.25$ & $ -0.03 \pm 0.18$ \\
TU CVn         &  HD112264      &  M5III         &  SRB           & $ 3345 \pm 30$ & $  0.49 \pm 0.17$ & $ -0.16 \pm 0.12$ \\
EK Boo         &  HD130144      &  M5III         &  LB:           & $ 3256 \pm 50$ & $  0.81 \pm 0.25$ & $ -0.24 \pm 0.18$ \\
RR UMi         &  HD132813      &  M5III         &  SRB           & $ 3387 \pm 39$ & $  0.46 \pm 0.25$ & $ -0.09 \pm 0.18$ \\
V1351 Cyg      &  HD186532      &  M5III         &  LB            & $ 3305 \pm 39$ & $  0.51 \pm 0.17$ & $ -0.08 \pm 0.12$ \\
30 Her         &  HD148783      &  M6III         &  SRB           & $ 3221 \pm 43$ & $  0.68 \pm 0.17$ & $ -0.24 \pm 0.12$ \\
AK Hya         &  HD073844      &  M6III         &  SRB           & $ 3130 \pm 58$ & $  0.68 \pm 0.17$ & $ -0.18 \pm 0.12$ \\
FH Vir         &  HD115322      &  M6III         &  SRB           & $ 3329 \pm 50$ & $  0.59 \pm 0.25$ & $ -0.16 \pm 0.18$ \\
V CVn          &  HD115898      &  M6IIIa        &  SR            & $ 3180 \pm 99$ & $  0.47 \pm 0.42$ & $ -0.32 \pm 0.30$ \\
RZ Ari         &  HD018191      &  M6III         &  SRB           & $ 3206 \pm 48$ & $  0.75 \pm 0.25$ & $ -0.19 \pm 0.18$ \\
omi Cet        &  HD014386      &  M7IIIe        &  M             & $ 2878 \pm108$ & $  0.41 \pm 0.76$ & $ -0.15 \pm 0.18$ \\
R Aqr          &  HD222800      &  M7IIIe        &  M+ZAND        & $ 3000 \pm118$ & $  0.26 \pm 0.42$ & $ -0.34 \pm 0.12$ \\
R Cas          &  HD224490      &  M7IIIe        &  M             & $ 2869 \pm159$ & $  0.17 \pm 0.76$ & $ -0.05 \pm 0.24$ \\
SW Vir         &  HD114961      &  M7III         &  SRB           & $ 2933 \pm 58$ & $  0.31 \pm 0.25$ & $ -0.32 \pm 0.06$ \\
R Leo          &  HD084748      &  M8III         &  M             & $ 2885 \pm 48$ & $  0.01 \pm 0.17$ & $ -0.39 \pm 0.06$ \\

\hline
\end{tabular}
\end{center}
\tablefoot{
\tablefoottext{a}{Spectral classification is reported from SIMBAD.}
\tablefoottext{b}{Variability types are taken from the General catalogue of Variable Stars and their description is as follows: E stands for eclipsing binaries, M for Mira type variables, SR for semi-regular variables, SRB for semi-regular, late-type giants, LB for slow irregular variables of late spectral type, and ZAND for symbiotic variables of the Z Andromedae type. NSV? indicates that the star is a member of New catalogue of Suspected Variable, but with unknown variability type. Colon denotes the uncertainty on the variability type.}
\tablefoottext{c}{The \feh{} measurements shall be regarded as uncertain, see Sect.~\ref{sec:mgiants}.}
}
\end{table*}

\subsection{Validation of the interpolator}\label{sec:validation_v2}

A potentially severe drawback in the approach presented in Sect.~\ref{sec:measV1} to assess the validity of the interpolator is that the interpolator depends on the spectra used afterwards to test it. In fact, if we had a pure interpolator that reproduces exactly the input spectra, the inversion would, by construction, restore the input parameters. In this case, the absence of difference between the input and inverted parameters would only demonstrate the self-consistency of the process. It would not tell anything about the correctness of the interpolated spectra.

Our polynomial interpolator, however, is smoothing the specific features of individual stars. A measure of the smoothing is given by the ratio between the degree of freedom of the interpolator (about two dozen) and the number of spectra in the library (close to a thousand). 
An actual spectrum from the library deviates from an idealized model in a number of ways: (i) it is affected by noise and residuals of instrumental signature, (ii) the input atmospheric parameters of the star have some uncertainty, and (iii) the star has a specific peculiarity (i.e. it is not fully described by the sole three parameters considered here). In the densely populated regions of the parameter space, these deviations are expected to be smeared out by the interpolator. In the margins, however, the interpolated spectra bear the signature of the individual spectra. In these cases, the consistency of the self-inversions does not probe the correctness of the interpolator.

To overcome this limitation, we computed a series of interpolators where each star was in turn dismissed (hereafter X interpolators). This provides us with the possibility of analysing each spectrum with an independent interpolator, and measuring by comparison the influence of any spectrum on the V2 interpolator. 


\section{Measurement of atmospheric parameters, biases, and errors}\label{sec:results}

\subsection{Atmospheric parameters}\label{sec:atm_params}

\begin{figure}[h]
  \centerline{\includegraphics[width=7.5cm,height=11.5cm]{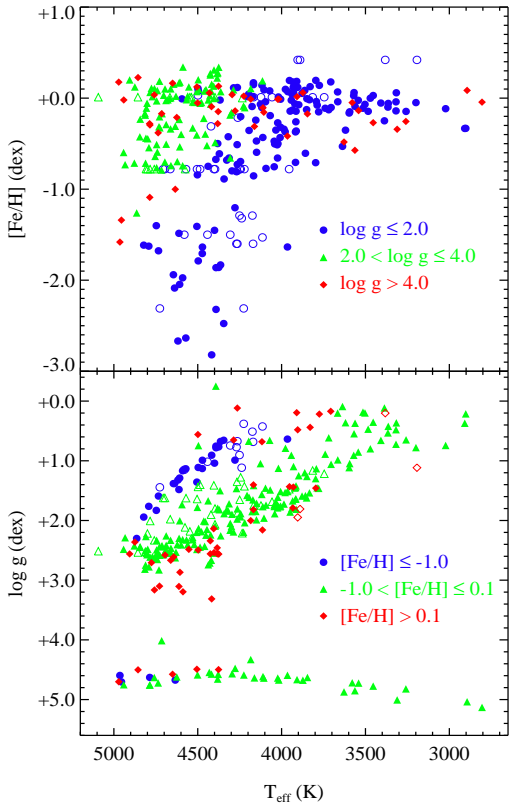}}
  \caption{Distributions of derived parameters
    in the \teff{}-\logg{} and \teff{}-\feh{}
    planes. Different classes of metallicity and gravity are shown with
    different colours, as indicated in the legend. Open symbols represent stars belonging to clusters.
  }
  \label{fig:distribution}
\end{figure}

Using the V2 interpolator and ULySS, we redetermined the atmospheric parameters of 331 stars. 
The spectra of cluster stars were fitted with \feh{} tight to the cluster's metallicity. The metallicity determination of these stars is presented and discussed in Sect.~\ref{sec:clusters}.
The results are listed in Table~\ref{tab:params} and Fig.~\ref{fig:distribution} presents the distribution of the stars in the parameter space.
The parameters measured with V2 are compared to the compilation in Fig.~\ref{fig:biases} (right column, the cluster stars are not shown on the \feh{} panel), and the corresponding statistics are reported in Table~\ref{tab:biases}.

We also separately fitted the blue (3600 -- 5500\,\AA) and the red (5600 -- 7400\,\AA) segments of the spectra, allowing us to detect peculiarities.
For example, the hotter star of a binary system would contribute more to the blue segment than to the red, leading to different solutions.
In Table~\ref{tab:params}, we assigned a quality "0" to the measurements when the \teff{} solution is consistent within 50~K (239 stars), "1", when the difference is within 50 to 100~K (51 stars), and "2" (41 stars) when it is larger.

The MILES spectra are assemblages of blue and red observations
connected in the region 5000 -- 5630\,\AA, with a third observation in the whole wavelength range, at a lower dispersion and with a wide slit, for the purpose of flux calibration. Therefore, three pointings on different nights were required for each final spectrum, multiplying the risk of misidentification. A discrepancy between the blue and red solutions is a diagnostic for this kind of a problem. 
The following two MILES spectra are likely affected by confusion of the target:
\begin{description}
\item[MILES~501:] The target was Arcturus (HD\,124897), a K1.5~III star whose \teff{} is expected to be around 4300\,K. Our inversions of the blue segment give the solution [$5301, 4.17, -0.14$] (in the rest of the paper we 
adopt this short bracketed description of the parameters) and that of the red segment [$4245, 1.94, -0.70$].
It is very likely that a pointing error resulted in using a wrong target 
(a G8V star) for the blue segment. The flux calibration was made with Arcturus.
We report the solution for the red segment (5600\,-\,7400\,\AA) in Table~\ref{tab:params}. 

\item[MILES~952:] The target was a star in the MESSIER\,71 globular cluster.
The blue segment corresponds to a 4100\,K bright giant, and the red, having a lower S/N to a 4900\,K subgiant star. 
The spectrum was flux calibrated using the 4850\,K star.
Considering the \teff{} adopted in Cenarro (4883\,K), we believe that the red segment corresponds to the target, and the blue segment to another star.
Similar values of the metallicities in the two domains suggest that both stars are in the same cluster. 
We discarded this star from our selection because its compiled value is at the hot limit of our sample.

\end{description}

These two spectra are of course discarded when computing the interpolator.

Another spectrum that attracted our attention is also likely a misidentification.
The target for MILES~591 was Gl617B (named HD\,147379B in MILES), a M3V star, for which the parameters 
in \citet{cenarro2007} match well with those of \citet{neves2013}.
However, as already noted by \citet{Prisinzano2012}, the spectrum resembles more a M0 or a M1 star than a M3 star.
This strongly suggests that the observed star is in fact HD\,147379 = Gl617A, the brightest component of the pair (two magnitudes brighter and located at about one arcmin). The fitted parameters would then agree with those derived by \citet{neves2013} and with the spectral type. HD\,147379 is also in CFLIB \citep{valdes2004}.
We changed its identification in Table~\ref{tab:params}.

The biases on \teff{} (Fig.~\ref{fig:biases}, right column, top panel) are now insignificant over the whole temperature range.

The biases on \logg{} have been marginally reduced. The measured gravities of the giants display a concentration around  \logg{} $\simeq 2.6$\,dex made of warm stars (\teff{}\,$\geq$\,4600\,K).
A weakly marked concentration, near \logg{}\,$\simeq$\,1.60\,dex, contains cooler stars (3800\,<\,\teff{}\,$\leq$\,4300\,K). In the first concentration, the dispersion of measured \logg{} is about half of that in the literature.
The intrinsic limited precision of spectroscopic gravities, affecting our compilation, did not allow us to investigate this feature further.

The \feh{} biases have also been generally reduced. At \feh{} $\le-1.0$\,dex, the metallicity bias of the giants is $-0.01$ dex, which is not significant. By comparison, the V1 interpolator gave a bias of $0.13$\,dex.
The metallicity of the cool dwarfs (\teff{} $\le 4000$~K) is properly determined, while the PVK and \citet{wu2011b} values were severely biased.

\subsection{Stability of the interpolator}\label{sec:validation_assess}

In addition to the above measurements obtained by the self-inversion of V2, we repeated the
analysis with the X interpolators described in Sect.~\ref{sec:validation_v2}. We obtained this second set of atmospheric parameters with exactly the same fitting parameters, and this set is not affected by the fact that the analysed spectra were used to make the interpolator. Therefore, comparing the two series allows us to measure the influence of any individual star on the interpolator.

As the ratio between the number of stars in the library and the degree of freedom of the interpolator suggested, this effect is generally very small.
For 200 of the 331 stars (about 60\%), the difference between the V2 and X solutions is smaller that 3~K on \teff, and 0.01~dex on both \logg{} and \feh. 
For only 25 stars (less than 10\%), this difference amounts to more than 30 K on \teff{} or 0.1 dex on the two other parameters. They are all at the margin of the parameter space, either with very low metallicity or at contrary super-solar, with the lowest or highest gravity, and generally among the coolest stars.

A close examination of the cases where the difference is large allows us to sort them into two categories: (i) some individual stars significantly constrain the interpolator, or (ii) some fits are unstable. In the first category, the X solution (i.e. with an interpolator that does not include the considered star) is usually moved towards the core of the parameter space because the interpolator becomes more noisy at the locus of the input atmospheric parameters. By comparison, the V2 solution is closer to the input parameters. The second category corresponds to either poor fits (due to peculiarities or low S/N) or to regions of the parameter space where the solution is not well defined. The latter occurs when the solution is found in an elongated $\chi^2$ valley with multiple local minima, or with a nearly flat bottom. In this case, changing the parameters of the fit slightly, such as the degree of the multiplicative polynomial or the wavelength range, or switching between the V2 and an X interpolator, may significantly affect the solution. 

On the basis of the comparison between the V2 and X series, we have defined a stability flag that is reported in Table~\ref{tab:params}. The flag is "0" if the difference between the derived parameters is less than 3\,K on \teff{} and 0.01 dex on both \logg{} and \feh{} (200 stars), "2" if the difference is more than 30\,K on \teff{} or 0.1\,dex on any of the two other parameters (25 stars). In rest of the cases (106 stars), the flag takes the value "1".

\subsection{Error analysis}\label{sec:error_analysis}

The fitting procedure produces error estimates (hereafter, fitting errors) that reflect the effect of the noise attached to the data.
Since the noise is not provided with the library spectra, we assume that the fitting residuals are pure noise, i.e. the fit is perfect and therefore $\chi^2 = 1$. Doing so, the estimated fitting errors are upper limits to
the errors due to the noise. On \teff{}, this error is of the order of 7 K.
It does not account for uncertainties attached to the 
interpolated spectra. The external error 
can be estimated by comparing our solution to the
compilation. If we conservatively assume that both series of measurements have 
the same precision, the errors are $1/\sqrt{2}$ times the standard deviations
of the two series. The inferred precision on \teff{} is about 60 K.
The same contrast between the fitting and external errors exists for other parameters. 

For an independent confirmation of the very small magnitude of the fitting errors,
we analyse a series of spectra of a given star. 
A good place to find this kind of series is the telescope archives of planet-hunting programmes.
Indeed, these programmes use long series of high-resolution
observations of their target sample to detect the velocity variations that
mark the presence of planetary companions.
We, therefore, downloaded 43 spectra of Gl1 (M1.5V) observed with HARPS, 
an echelle spectrograph attached to the ESO 3.6m telescope. These
observations were acquired along the GTO programme 072.C-0488(E) and we used the
data reduced by the automatic pipeline. Our analysis with ULySS and the
new interpolator is insensitive to velocity shifts and flux calibration,
and the spectra essentially differ only by their noise.
We matched the resolution to MILES by convolving the spectra with a Gaussian of
velocity dispersion of 70 \kms, and used the wavelength range 5400 to 6500 \AA.
For these 43 spectra, we found the mean parameters [3377, 4.63, -0.22]. 
The mean internal errors are 6.5, 0.024, and 0.025 for the three 
parameters, respectively, and the standard deviation of the solution 5.0, 0.024, and 0.010, respectively.
The standard deviations are comparable to the mean fitting errors, confirming
that the latter give the proper estimate of the effect of the noise.

A simple test establishes that the total errors, including all the sources of uncertainty, are in fact larger than the fitting errors. We analysed the blue and red segments of the MILES spectra separately and compared the solutions.
For the 40 stars belonging to the first class reported in Table~\ref{tab:biases},
the standard deviation between the two series of \teff{} measurements 
is about 39\,K while the estimated external error is about 58\,K for these stars.
Similarly for \logg{} and \feh{}, the values of standard deviations between the two series
are $\approx$~0.21\,dex and 0.14 dex, respectively, while the mean external
errors on these parameters are 0.27\,dex and 0.16 dex, respectively.
The biases and dispersions of the two series of parameters with respect to the literature are comparable, and there is no indication that some particular wavelength region would constrain one parameter or another better.
This analysis shows that the dispersions between the two series are considerably larger than the fitting errors, and that they are approaching the external errors.
We can list various reasons why the external errors are larger than the fitting errors:
(i) The stellar spectra cannot be perfectly modelled from the three atmospheric
parameters. Other characteristics, like binarity, rotation, chromospheric activity
or detailed abundances can explain some mismatch of the spectra. 
(ii) The uncertainities in the atmospheric parameters of the stars used to build the interpolator propagate to the interpolated spectra. 
(iii) The analytical form of the interpolator does not perfectly reproduce the spectra. 

In \citet{wu2011b}, the total errors were evaluated by scaling up the fitting errors. It was found that both errors were roughly proportional to each other, and, therefore, scale factors were computed by comparing the fitting errors to the external errors derived from comparisons to the literature. 
Following the same approach, we calculated the scaling factors on the whole sample. 
We found that the relative error on \teff{} and \logg{} has to be rescaled by a factor nine, and the errors on \feh{} by a factor eight.
In order to compare our estimated errors with those in PVK,
we computed the ratio between the two errors for the coolest stars (\teff{}$\leq$\,3800\,K) and for the others (\teff{}\,>\,3800\,K). We found these ratios to be close to one for all three parameters and the two temperature regimes, which implies that the two error estimates are almost equivalent.

\subsection{Remarkable stars}\label{sec:outliers}

We carefully examined 
the remarkable stars with extreme locations in the three parameters.
Since discrepancies can result from either inaccurate values in the compilation
or wrong solution of the fit or both, we rechecked the
literature and the quality of our fit, and proceeded to appropriate corrections if needed. 
After each correction, the whole procedure was repeated, and the stars listed in Table~\ref{tab:out} or labelled in Fig.~\ref{fig:biases} are 
those that resisted our attempts to resolve them. In this section, we discuss them individually.

To check the quality of the fits for those spectra, we used four diagnostics. 
First we examined the residuals closely (difference between the observed 
spectrum and the fitted spectrum) to search for discrepant spectral features 
that would be reminiscent from some peculiarity of the star. Second, we 
compared the parameters derived from fitting the red and the 
blue regions of the spectrum separately and masking the regions of possible coronal 
emission. This test could also reveal some peculiarities, particularly 
binarity (if the two components have different spectral types). Third, we 
examined the  $\chi^2$ maps and, finally, we searched the literature exhaustively for 
possible explanations of the discrepancy.

\begin{figure*}
     \centering
     \begin{subfigure}[b]{0.235\textwidth}
       \centering
       \includegraphics[width=\textwidth]{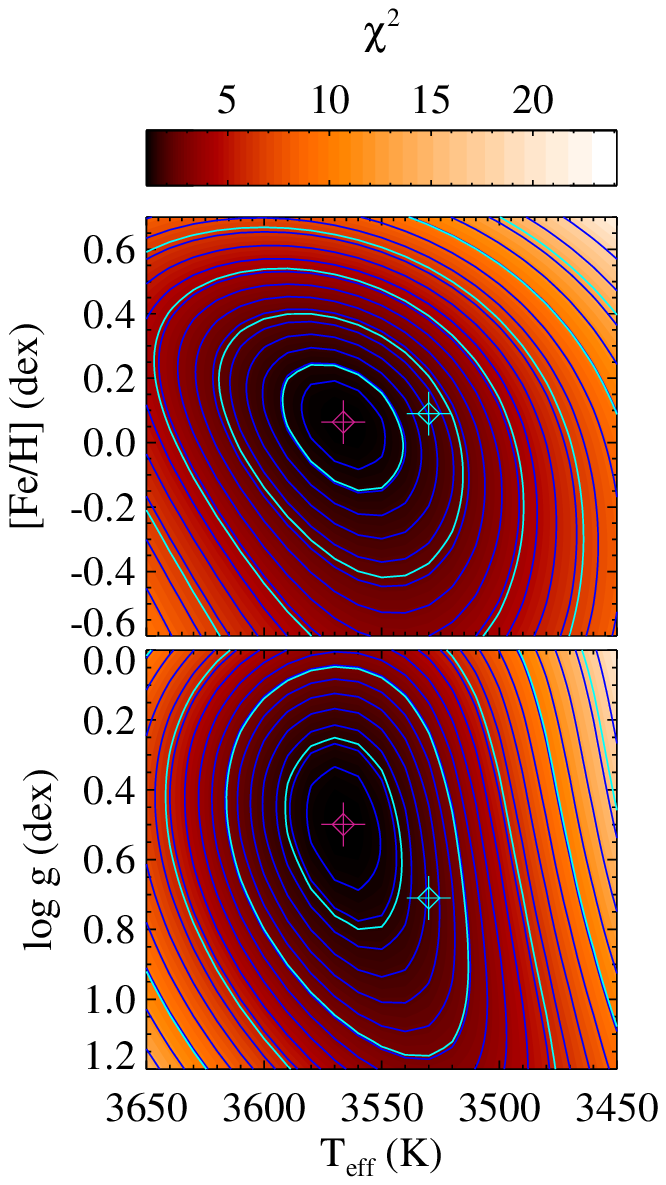}       
     \end{subfigure}%
     ~ 
     \begin{subfigure}[b]{0.235\textwidth}
       \centering
       \includegraphics[width=\textwidth]{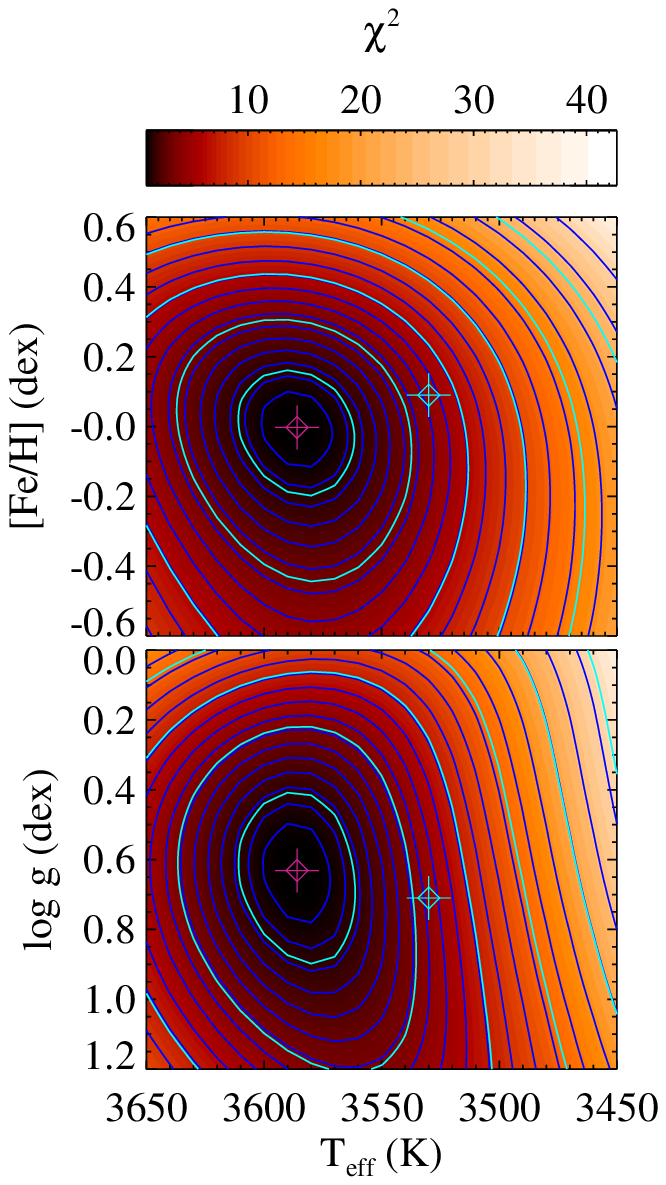}
     \end{subfigure}
     ~ 
     \begin{subfigure}[b]{0.235\textwidth}
       \centering
       \includegraphics[width=\textwidth]{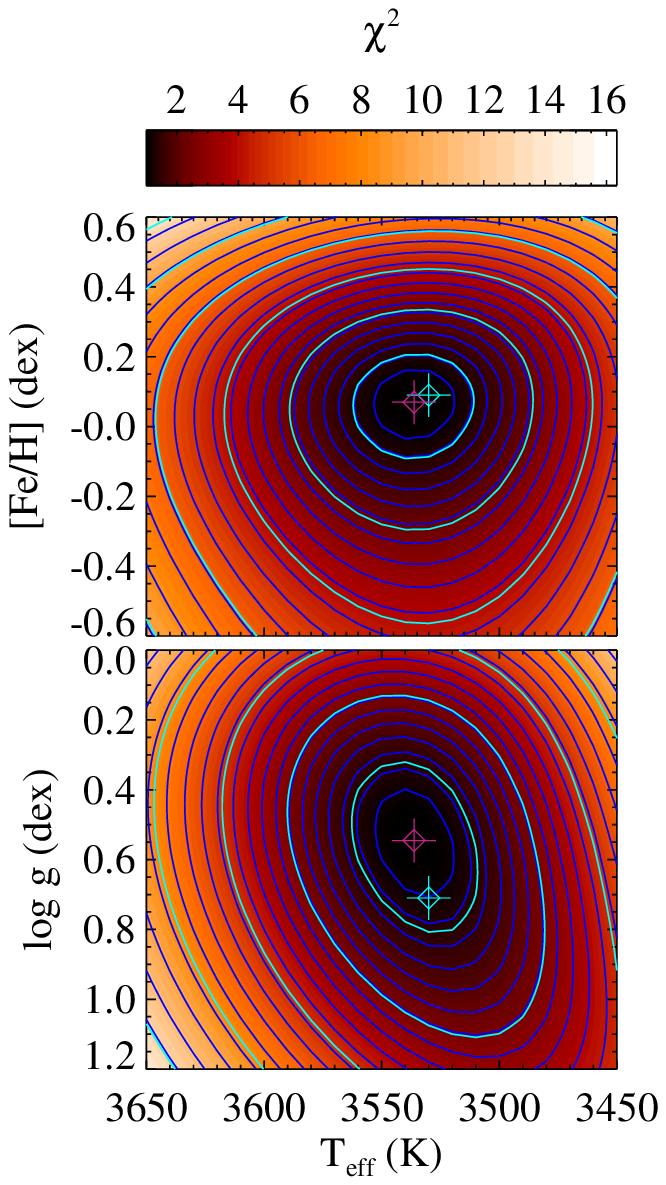}
     \end{subfigure}
          
     \caption{$\chi^2$ maps for HD\,61913
       with different fitting wavelength ranges.
       First with the complete wavelength range used (3600~--7400~\AA{}), second
       3600~--~5500~\AA, and third 5500~--~7200~\AA. 
       The red crosses indicate the $\chi^2$ minimum, and the cyan crosses the compiled parameters.
       The inner contour in cyan correspond to the external errors, and the other contours correspond to the 2, 3, 4 ... $\sigma$ levels.}\label{fig:chimaps1}
   \end{figure*}

\begin{figure}
  \centerline{\includegraphics[width=7cm,height=11.5cm]{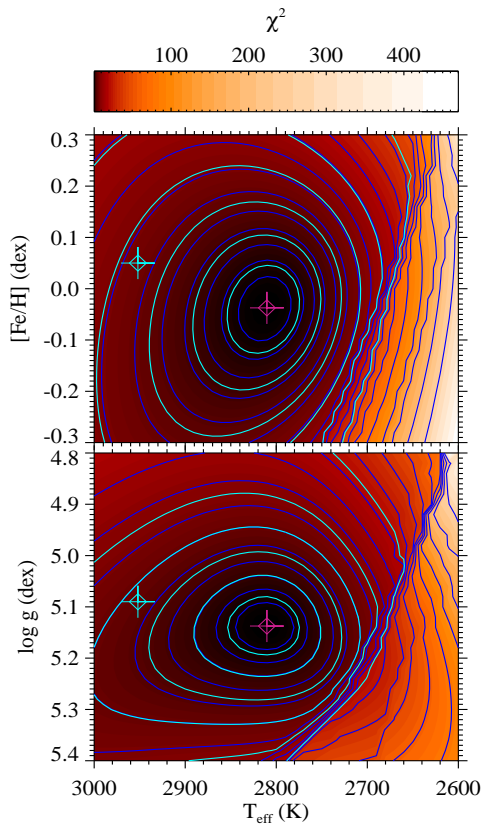}}
  \caption{$\chi^2$ map for G156-031 showing the dissymmetry at the edge of the parameter space.
  }
  \label{fig:chi_disc}
\end{figure}

The $\chi^2$ maps are used to explore the topology of the parameter space. 
They reveal the degeneracies between the parameters and possible anomalies 
resulting in an ill-defined minimum region (or presence of local minima). 
The maps were computed by cutting the parameter space in two 
planes: the \teff{} vs. \feh{} and \teff{} vs. \logg{}, 
fixing \logg{} and \feh{}, to their values at the adopted solution, respectively.
Figure~\ref{fig:chimaps1} 
shows three typical maps for a $\approx$3500~K star, where the fit is 
performed on the whole spectral range and also separately on the blue  
([3600, 5500] \AA{}) and red  ([5500, 7200]~\AA{}) regions. The cyan contours 
overlaid correspond to 1 to 5-$\sigma$.
The three maps are similar, and the location of the solution is consistent 
within 1-$\sigma$. The elongated shape of the contours, in particular, on the 
\teff{} vs. \feh{} plane reflects the degeneracy between the parameters.
For good data and normal stars, the maps as well as the parameters do not depend on the fitted wavelength range.
Figure \ref{fig:chi_disc} shows the $\chi^2$ maps for G 156-031, a star at the edge of the parameter space populated by the MILES stars. 
The $\chi^2$ values increase sharply on the low \teff{} side, resulting in a dissymmetry that is potentially at the origin of a bias.
We discuss below the stars labelled in Fig.~\ref{fig:biases} and listed in Table~\ref{tab:out}.

\begin{table}
\centering
\caption{Remarkable stars labelled in Fig.~\ref{fig:biases}}
\begin{tabular}{llcl}
\hline\hline
No & Name & MILES \\
\hline
 1& V1855 Ori = MS 0515.4-0710 & 175 \\
 2& HD\,113285                   & 459 \\
 3& HD\,126327                   & 508 \\
 4& MS 1558.4-2232             & 580 \\
 5& HD147923 ?                 & 593 \\
 6& G156-031                   & 838 \\
 7& BD+19 5116B                & 884 \\
 8& G171-010                   & 890 \\
 9& Cl*NGC5904 ARP II-51       & 927 \\
10& NGC6838 1053               & 964 \\
11& NGC7789 329                & 972 \\
12& NGC7789 637                & 980 \\
13& HD\,44889                    & 220 \\
14& HD\,56577                    & 250 \\
15& HD\,96360                    & 400 \\
16& HD\,200905                   & 784 \\
17& Cl*NGC6791 GVZH R5         & 941 \\
\hline
\end{tabular}
\label{tab:out}
\end{table}

(1) V1855 Ori = MS 0515.4-0710 (MILES~175) is a K0 or K2 type star. 
V1855 Ori and MS 0515.4-0710 currently appear as two distinct sources in SIMBAD, but their cross-identification with a unique source leaves no doubt, and allows us to complete our compilation.
\citet{favata1997} and \citet{cenarro2007} determined [4570, 3.50, 0.16] using the Str\"omgren photometry of \citet{morales1996} and the calibration relation of \citet{alonso1996}. 
\citet{xing2010} gives \teff{}\,=\,4908\,K and \logg{}\,=\,4.45\,dex, computed from the $(B-V)$ colour and calibrations of \citet{casagrande2006}.
\citet{biazzo2012} give [5100, 4.2, 0.05], based on high-resolution spectroscopic observations. 
\citet{martin1994} adopted \teff{}\,=\,5150\,K from their own relation between the spectral type and the temperature.
In our compilation we adopt the average parameters [4860, 4.05, 0.10].
The parameters obtained after fitting the blue range, 
[$5382, 4.52, 0.16$], and the red range, [$5179, 4.42, 0.01$], differ significantly, as is often the case for variable stars. 
Our adopted measurement is not discrepant with the analysis of \citet{biazzo2012} and therefore we trust our measurements.

(2) HD\,113285 (MILES~459) is a M8 III type pulsating variable star. This star belongs to the LICK/IDS library \citep{W94}. \citet{gorgas1999} determined \teff{}\,=\,2924\,K by extrapolating the \teff{} versus spectral type relation of \citet{ridgway1980}. \citet{gorlova2003} obtained \teff{}\,=\,$2900\pm320$\,K. The parameter \teff{} was derived from the spectral type and \logg{} was obtained from \teff{}, luminosity, and mass. \citet{mcdonald2012} derived \teff{}\,=\,2602\,K using SED fitting, neglecting Galactic extinction and assuming a solar composition. The authors PVK adopted [$2924, 1.50, -$], and in our compilation we adopt the average of the available measurements [$2900, 0.00, -$], where \logg\,=\,0.0 dex is a reasonable guess for this type of star, and is likely to be accurate within $\pm0.50$ dex. The Fe content has not been measured.
With V2 we determine [$2902, 0.21, -0.33$]. This star is one of the two coolest giants of the sample and because of the lack of \feh{} references in this region of the parameter space, our determination of the metallicity cannot be trusted.

(3) HD\,126327 (MILES~508) is a M7.5-M8 pulsating star \citep{tsuji2008}, which is pretty similar to HD\,113285 discussed above. \citet{perrin1998} report photometric \teff{}\,=\,$2786\pm46$\,K. \citet{dyck1998} obtained \teff{}\,=\,$2915\pm113$\,K using the infrared flux method (IRFM). With the same method, \citet{tsuji2008} determined \teff{}\,=\,2850\,K. \citet{mcdonald2012} determined \teff{}\,=\,2581\,K using SED fitting. 
\citet{gorgas1999} and \citet{cenarro2007} adopted a metallicity of $-0.58$ dex, which is in fact a measurement of [C/H] by \citet{tsuji1986}.
In PVK, the adopted parameters [$3000, -, -0.58$] are taken from \citet{cenarro2007}.
In our compilation, we adopt [$2850, 0.00, -0.58$] where \teff{} is an average of the available determinations. The value \logg{}$\,=\,0.0$ dex is a reasonable guess for this type of star and we kept the metallicity of \citet{tsuji1986}, knowing it is not a reliable estimate of \feh.
Using V2 we determine [$2908, 0.37, -0.33$]. The value of the metallicity may not be considered as reliable for this star also.

(4) MS 1558.4-2232 (MILES~580, K1 IV) is a variable star of BY Draconis type \citep{kaza2001} whose variability is due to rotation coupled with starspots and chromospheric activity. It is a member of the Scorpius A association.
\citet{favata1997} carried out a spectroscopic analysis resulting in [$4250, 3.50, 0.10$], where the \teff{} is based on Str\"omgren photometry, and \logg{} is adopted as a typical value for a K giant. This was later adopted by \citet{cenarro2007} in their compilation. We did not find any other measurement in the literature. 
The parameters determined in PVK are [$4727, 4.02, -0.14$], while using V2 we obtain [$4715, 4.01, -0.17$], which is stable over the whole wavelength range.
As the fit of the spectrum is good, we trust our result.
In general, the residuals for the K dwarfs are large, possibly because of
imprecise measurements in the literature.

(5) For HD\,147923 (MILES~593: M2), \citet{smith1990} derived [$3600, 0.80, -0.19$], comparing the equivalent width of observed lines in near-infrared high-resolution spectra with the predicted LTE equivalent widths of synthetic spectra computed from model
atmospheres. \citet{cenarro2007} adopted these parameters from \citet{smith1990} in their compilation.
\citet{mcdonald2012} determined \teff{}\,=\,3800\,K. 
Values determined in PVK, [$4773, 4.69, -0.26$], are consistent with the present determination (using V2) of [$4787, 4.76, -0.28$].
The spectrum is well fitted and the $\chi^2$ map has regular contours and, 
therefore, we trust our measurements.
The considerable discrepancy between the compilation and our measurements suggests a misidentification. The star is pretty isolated on the sky and we do not think it can be a matter of pointing accuracy.
Since we are not able to correct the designation, we kept the original identification in Table~\ref{tab:params}, but attached a "?" to mark our suspicion. 

(6) G\,156-031 (M6 V, MILES~838) is the coolest star of the sample and is a part of a triple star system. 
It belongs to the Lick/IDS library \citep{W94} and \citet{gorgas93} obtained \logg{}\,=\,5.09\,dex. \citet{cenarro2007} adopted [$2747, 5.09, -$].
Using empirical spectral type standards and synthetic models, \citet{rojas2012}
calibrated the $\rm{H_2O}$-K2 index, which measures the absorption due to $\rm{H_2O}$ opacity, as an indicator of an M dwarf spectral type and effective temperature, and proposed \teff\,=\,$2952\pm23$ K and a M5 type. They estimated \feh{}\,=\,$0.05\pm0.17$ dex from measurements of Na I, Ca I, and $\rm{H_2O}$-K2. 
\citet{casagrande2008} determined
\teff{}\,=\,2650\,K for this star using flux ratio in different bands, but
with questionable accuracy owing to the fact that the star lies in a triple system.
\citet{dawson2000} obtained \teff{}\,=\,3000\,K, \logg{}\,=\,5.0\,dex, and \rm{[M/H]}\,=0.0\,dex by comparing spectroscopic observations with synthetic spectrum.
In our compilation, we adopt [$2952, 5.09, 0.05$] for this star.
The \teff{} and \feh{} are taken from \citet{rojas2012}, which are most recent determinations, and \logg{} is taken from \citet{gorgas93}.
With V1 interpolator, the \teff{} was about 300 K warmer with \feh{}\,=\,$-1.28$\,dex; this was a highly biased determination at the edge of the parameter space.
Because the blue region of the spectrum has a significantly lower S/N, 
we performed the fit with the V2 interpolator in the range [5000,7200~\AA{}], leading to [$2805, 5.13, -0.04$].
Our \teff{} estimate is about 100 K cooler than that determined in \citet{rojas2012}, but is within the range of the other estimates. 
In PVK, this star was a prominent outlier, and with V2, the determinations have clearly improved and are now consistent with the compilation.

(7) BD+19\,5116B (Gl\,896B, MILES~884) is a M4.5 flaring dwarf star with \feh{}\,=\,0.14 dex \citep{bonfils2005}.
\citet{morales08} assessed \teff{}\,=\,3080\,K through infrared colour and spectral index. 
\citet{cenarro2007} compiled [2950, 5.06, 0.10], but the source is not explicitly mentioned.
In our literature compilation, we adopt the parameters as [$3080, 5.06, 0.14$].

Fitting with V1 in the region [5000,7200~\AA{}] yields [$3454, 4.78, -1.07$], which is illustrative of the well-identified bias on the cool dwarfs. The temperature is overestimated, and the metallicity is underestimated.
With the use of V2 in the same wavelength region, we measure [$3259, 4.82, -0.26$].
The emission lines (mostly H \& K) are discarded with the automatic rejection of the spikes. 
Although the biases are well reduced, \teff{} is still $\sim$\,150\,K warmer than the compiled value and \feh{} remains subsolar, but with a large error bar. This is the only cool dwarf for which the new interpolator fails to match the literature values. Since our tests in Sect.~\ref{sec:extensioncooldwarfs} have established the reliability of our determinations in this regime, we trust our measurement.
   
(8) G\,171-010 (MILES~890, \citealp[M6.0 V;][]{jenkins2009}) is the second coolest star of MILES.
In our compilation, for this star \teff{} and \feh{} are $3058\pm65$\,K and $0.05$\,dex from \citet{rojas2010} and 
\citet{gorgas93}, respectively. \citet{jenkins2009} determined \teff{}\,=\,2746\,K
with the $V-K_{\rm{s}}$ relation taken from \citet{casagrande2008}. In \citet{cenarro2007} compiled parameters
are [$2799,5.12,-$] and PVK adopted the same.

Our fit yields [$2894, 5.04, 0.09$] with a regular $\chi^2$ map.
The temperature is about 150 K cooler than the compiled value and is between the two determinations of \citet{rojas2010} and \citet{jenkins2009}.

(9) Cl*NGC5904 ARP II-51 (MILES~927).
\citet{cenarro2007} compiled [$4627, 1.74, -1.11$], where \teff{} is from the $B-V$ and $V-K$ versus \teff{} relations of \citet{alonso1996,alonso1999}. \citet{carretta-paper7} report [$4587, 1.75, -1.33$] from high-resolution spectroscopy and
\citet[APOGEE]{meszaros2013} obtained \teff\,=\,$4809\pm134$\,K, \logg\,=\,2.15\,dex and [M/H]\,=\,$-1.25$\,dex.
In our compilation, we adopt [$4587, 1.75, -1.29$] where \feh{} is the cluster value from Table~\ref{tab:compil_globularclusters}.

PVK obtained [$6115, 3.84, -1.11$], and
using V2 we estimate [$5718, 1.98, -1.29$], where \feh{} is fixed at the cluster metallicity
(the star is not in the field of view of Fig.~\ref{fig:biases}).
This \teff{} estimate is about 1100 K warmer than the compiled temperature. Releasing the constraint on \feh{} gives [$5577, 1.83, -1.54$], where the metallicity is discrepant from the cluster value and conflicts with the membership ascertained from proper motion and radial velocity.
Fitting the spectrum in blue and red segments separately results in [$5622, 2.05, -1.29$] and [$5438, 2.45, -1.29$], respectively, where \feh{} is fixed, [$5520, 1.82, -1.51$], and [$5084, 1.71, -1.90$] with free metallicity.
The quality of these fits is poor, and because the star is in a crowded field, we can neither exclude a pointing error nor light contamination.
In any case, the MILES spectrum cannot be trusted.

(10) NGC6838 1053 (MILES~964) has been the object of a number of spectroscopic studies. In our compilation, we adopted [$4150, 1.50, -0.78$]
from \citet{melendez2009}. This is consistent with the values in \citet{cenarro2007}, [$4167, 1.51, -0.84$], and with a number of previous analysis listed in the Pastel catalogue. More recently, \citet{saviane2012} gave [$4659\pm89, 1.74\pm0.26, -0.84$].

The spectrum is well fitted, PVK derived [$4659, 1.74, -0.84$], and we determine [$4696, 1.62, -0.78$] using V2. The fitted \teff{} is significantly warmer than our compilation, but matches the Saviane et al. measurement.
A possible explanation could be the contamination by NGC6838~1052 located about 5 arcsec away.
The Fourth Naval Observatory Catalog, UCAC4 \citep{UCAC4} indicates that this star is about two magnitudes fainter, but is somewhat bluer ($J-K = 0.61$, vs. $0.94$ for 1053).

As a test, we analysed a spectrum taken with X-Shooter at the VLT in the frame of programme 189.B-0925(C) \citep{chen2014}, downloaded from the ESO archive. The blue and red segments were measured independently because they were observed with different arms of the spectrograph. Although the fit is considerably better for the blue segment, both give consistent results. The blue solution is [$4103, 1.45, -0.90$], in good agreement with our compilation, and hence supports the idea that the MILES spectrum is contaminated.

(11) NGC\,7789\,329 (MILES~972) is a K2 III type star \citep{barta2004}.
\citet{pila1985} obtained [$4330, 2.20, -0.10$], where \teff{} is determined from the $(B-V)$ colour index measured
by \citet{burbidge1958} and the $(B-V)$ vs. temperature calibration
of \citet{bohm1981}, and \feh{} is obtained from the high-resolution spectroscopic observations.
\citet{cenarro2007} report [$4527, 2.14, -0.13$], where \teff{} is derived from the $B-V$ vs. \teff{} relations of \citet{alonso1996,alonso1999}.
\citet{jacobson2011} estimated [$4500, 2.2, 0.05$], where \teff{} is the average
of the $(B-V)$, $(V-K)$, $(J-K)$ temperatures obtained using the calibrations of \citet{alonso1999}.
In our compilation we adopt the average parameters [$4500, 2.18, 0.01$].
This temperature is consistent with the spectral classification.

PVK determined [$5043, 2.30, -0.13$] and using V2, our determinations are [$5093, 2.52, 0.01$] by fixing \feh{} at cluster metallicity and with a fit of good quality. We have no explanation for the discrepancy.

(12) NGC7789\,637 (MILES~980, K2 III) is another \teff{} outlier.
\citet{cenarro2007} compiled [$4561, 2.24, -0.13$] from the $B-V$ versus \teff{} calibrations.
\citet{jacobson2011} report [$4500, 2.30, 0.05$], where \teff{} is derived using the \teff{}-colour calibrations of \citet{alonso1999}
and the $B-V$, $V-K$, and $J-K$ colours. 
In our compilation we adopt [$4500, 2.30, 0.01$].
PVK determined [$4812, 2.37, -0.13$], and using V2 we determine [$4857, 2.54, 0.01$] by fixing the \feh{}
at cluster metallicity value and obtaining a good fit.
No reason could be found for this discrepancy.

(13) HD\,44889 (MILES~220) is a K0 type star. 
\citet{cenarro2007} adopted [$3775, 0.40, -0.20$], which was derived by \citet{castilho2000}.
Different effective temperature values were derived from colours($B-V=1.71$ and $V=7.55$; obtained from CDS) based on the calibration tables of \citet{bessell1998} and \citet{lejeune1998} and then averaged to obtain the final \teff{}.
\citet{mcdonald2012} determined \teff{}\,=\,3875\,K using SED fitting. 
The K0 classification reported by \citet{cenarro2007} comes from the Hipparcos Input Catalogue and would correspond to \teff{}\,$\approx$\,4775\,K \citep[see Table 6 of][]{martin1994}, which is in marked disagreement with the other determinations. There is no detailed spectroscopic study of this star and the compiled parameters have low reliability.
In our compilation, we adopt \logg{} and \feh{} from \citet{castilho2000}, and for \teff{} the average of the two available determinations.
PVK measured [$4022, 1.44, -0.16$] and using the V2 interpolator we obtain [$4006, 1.51, -0.32$]. The estimated \logg{} value also departs significantly from the compiled value. The fit of this star 
is good and there are no large residuals seen for this star. Our \teff{} determination is only $\sim$ 100\,K hotter than that in \citet{mcdonald2012}. Therefore, we believe that our determinations are reliable.

(14) HD\,56577 (MILES~250) is a K4 III type star and is a part of a double star, together with HD\,56578, a A4m star lying 27 arcsec away.
\citet{luck1982} reports [$4500, 1.35, 0.15$] for this star. Recently \citet{mcdonald2012} determined \teff{}\,=\,3944\,K for this star using SED fitting.
The values of the parameters in \citet{cenarro2007} are [$4372, 1.25, -0.20$], where the sources are not explicitly cited. 
In our compilation, we adopt [$4158, 1.25, 0.15$], where \teff{} is the average from \citet{cenarro2007} and \citet{mcdonald2012}.
In PVK, the parameters are [$3944, 0.72, 0.14$] 
while using the V2 interpolator, we determine [$3904, 0.48, 0.12$],
consistent with the spectral classification and with \citet{mcdonald2012}. 
The good quality of our fit allows us to rule out the higher \teff{} from earlier studies.

(15) HD\,96360 (MILES~400). We compiled [$3550, 0.50, -0.58$], following \citet{smith1990} and \citet{cenarro2007}.
We did not find detailed spectroscopic studies in the literature.
Our measurement, [$3471, 0.80, 0.00$], significantly departs in \feh. Although the compiled value is itself uncertain, we should stress that metallicity measurements are generally difficult in this region of the parameter space, and that we shall not grant them a high confidence.

(16) For HD\,200905 (MILES~784),
we compile [$4005, 1.21, -0.34$] from \citet{MW1990} and \citet{hekker2007}
and we measure [$3977, 0.79, 0.10$].
The high metallicity disagrees with our compilation but some older measurements cited in Pastel catalogue 
are consistent with a solar metallicity.
As a test, we analysed two spectra downloaded from the ELODIE archive. The data have a high S/N, and the two fits consistently return [$4037, 0.86, 0.09$]. The good agreement with our result rules out any anomaly of the MILES spectrum. 
The other stars in the same region of the parameter space do not share a similar bias and, therefore, we trust our determination.

(17) Cl*NGC6791 GVZH R5 (MILES~941) is the coolest red giant candidate detected in the surveys of NGC~6791 by \citet{garnavich1994} and \citet{stetson2003}. NGC~6791 is one of the more metal-rich open clusters.
From its radial velocity, \citet{garnavich1994} suggest the star belongs to the cluster and since the radial velocity of the cluster is different from the typical value in the field, the assessment is robust. In addition \citet{dias2014} estimate, using proper motions, a membership probability of 89\%. Therefore, the membership appears well established.
\citet{cenarro2007} derived \teff{}$ = 4057$  and \logg{}$ = 2.32$ by interpolating the
\citet{girardi2000} isochrones in B-V and M$_V$. However, 
we took \teff\,=\,3282\,K from \citet{buzzoni2010}, 
who used the colour-\teff{} calibrations of \citet{alonso1999} to derive \teff{} from $V-K$ and $J-K$. However, the high metallicity of the cluster results in a strong blanketing that makes the star appear redder at optical and NIR wavelengths. This decreases the reliability of the \teff{} estimates.
Our fit returned [$3194, 0.79, 0.03$], leaving the metallicity free, but the metallicity measurement is not reliable, as stated in Sect.~\ref{sec:mgiants}.

\begin{longtab}
\begin{center}
       \setlength{\tabcolsep}{7.0pt}
\begin{longtable}{llrrrl}
\caption{Super-solar giants}\label{tab:supersolar}
\\
\hline\hline

Name & MILES & \teff{} & \logg{} & \feh{} & Ref. \\

\hline
\endfirsthead
       \caption{continued.}\\
\hline\hline
Name & MILES & \teff{} & \logg{} & \feh{} & Ref. \\
\hline
        \endhead
       \endfoot

HD120933 & 487 & 3820 & 1.52 &$ 0.50 $  & 1\\
HD120933 &     & 3820 & 1.52 &$ 0.28 $ & 5\\
HD120933 &     & 3646 & 1.52 &$ 0.10 $ & C\\
HD120933 &     & 3529 & 0.98 &$-0.09$  & PVK\\
HD120933 &     & 3594 & 1.05 &$-0.15$  & M\\
\hline 

HD146051 & 590 & 3793 & 1.40 &$ 0.32$  & 1\\
HD146051 &     & 3679 & 1.40 &$ 0.32$  & 2\\
HD146051 &     & 3850 & 1.20 &$ 0.00$  & 11\\
HD146051 &     & 3721 & 1.02 &$-0.24$  & 12\\
HD146051 &     & 3779 & 1.20 &$ 0.00$  & C\\
HD146051 &     & 3783 & 1.45 &$-0.03$  & PVK\\
HD146051 &     & 3779 & 1.46 &$-0.15$  & M\\
\hline 

HD013520 & 080 & 3966 & 1.80 &$ 0.30$  & 1\\
HD013520 &     & 3970 & 1.70 &$-0.25$  & 5\\
HD013520 &     & 3900 & 1.60 &$ 0.30$  & 4\\
HD013520 &     & 3970 & 1.70 &$-0.24$  & C\\
HD013520 &     & 4043 & 1.66 &$-0.16$  & PVK\\
HD013520 &     & 4023 & 1.61 &$-0.27$  & M\\
\hline 

HD119228 & 481 & 3661 & 1.80 &$ 0.30$  & 1\\
HD119228 &     & 3600 & 1.60 &$ 0.30$  & 4\\
HD119228 &     & 3600 & 1.60 &$ 0.30$  & C\\
HD119228 &     & 3684 & 1.02 &$ 0.00$  & PVK\\
HD119228 &     & 3705 & 1.06 &$-0.10$  & M\\
\hline 

HD149009 & 603 & 3910 & 1.60 &$ 0.30 $  & 1\\
HD149009 &     & 3910 & 1.60 &$ 0.18 $  & 3\\
HD149009 &     & 3910 & 1.60 &$ 0.18 $  & C\\
HD149009 &     & 3877 & 1.23 &$ 0.12 $  & PVK\\
HD149009 &     & 3862 & 1.20 &$ 0.09 $  & M\\
\hline 

HD219734 & 871 & 3730 & 0.90 &$ 0.27$  & 1\\
HD219734 &     & 3730 & 0.90 &         & 6\\
HD219734 &     & 3730 & 0.90 &$ 0.27$  & 8\\
HD219734 &     & 3800 & 1.20 &$-0.04$  & 9\\
HD219734 &     & 3765 & 1.05 &$ 0.12$  & C\\
HD219734 &     & 3616 & 1.00 &$ 0.04$  & PVK\\
HD219734 &     & 3665 & 0.99 &$-0.04$  & M\\
\hline

HD138481 & 560 & 3910 & 1.46 &$ 0.25$  & 1\\
HD138481 &     & 3890 & 1.64 &$ 0.13$  & 5\\
HD138481 &     & 3915 & 1.46 &$ 0.13$  & C\\
HD138481 &     & 3917 & 1.15 &$ 0.00$  & PVK\\
HD138481 &     & 3898 & 1.25 &$-0.07$  & M\\
\hline 

HD061603 & 269 & 3870 & 1.50 &$ 0.24$  & 1\\
HD061603 &     & 3870 & 1.50 &$ 0.12$  & 3\\
HD061603 &     & 3870 & 1.50 &$ 0.12$  & C\\
HD061603 &     & 3983 & 1.41 &$ 0.22$  & PVK\\
HD061603 &     & 3953 & 1.43 &$ 0.19$  & M\\
\hline 

HD131918 & 530 & 3970 & 1.49 &$ 0.22$  & 1\\
HD131918 &     & 3970 & 1.72 &$ 0.15$  & 5\\
HD131918 &     & 4140 & 1.65 &$-0.24$  & 10\\
HD131918 &     & 4055 & 1.68 &$-0.04$  & C\\
HD131918 &     & 4154 & 1.65 &$ 0.00$  & PVK\\
HD131918 &     & 4118 & 1.56 &$-0.10$  & M\\
\hline 

HD175865 & 705 & 3420 & 0.50 &$ 0.14$  & 1\\
HD175865 &     & 3420 & 0.50 &         & 6\\
HD175865 &     & 3759 & 0.50 &$ 0.14$  & 8\\
HD175865 &     & 3297 & 0.50 &$-0.25$  & C\\
HD175865 &     & 3181 & 0.47 &$-0.29$  & PVK\\
HD175865 &     & 3316 & 0.36 & $0.06$  & M\\
\hline 

HD060522 & 264 & 3899 & 1.20 &$ 0.12$  & 1\\
HD060522 &     & 3902 & 1.20 &$ 0.12$  & 6\\
HD060522 &     & 4130 & 1.90 &$-0.36$  & 10\\
HD060522 &     & 3884 & 1.20 &$ 0.00$  & C\\
HD060522 &     & 3846 & 1.69 &$ 0.04$  & PVK\\
HD060522 &     & 3834 & 1.54 &$-0.02$  & M\\
\hline 

HD148513 & 597 & 3997 & 1.67 &$ 0.11$  & 1\\
HD148513 &     & 4000 & 1.77 &$ 0.00$  & 5\\
HD148513 &     & 4046 & 1.00 &$ 0.20$  & 6\\
HD148513 &     & 4075 & 0.30 &$-0.31$  & 7\\
HD148513 &     & 4200 & 2.05 &$-0.02$  & 10\\
HD148513 &     & 4044 & 1.67 &$-0.02$  & C\\
HD148513 &     & 4131 & 2.15 &$ 0.20$  & PVK\\
HD148513 &     & 4114 & 2.16 &$ 0.19$  & M\\
\hline 

\hline 
\end{longtable}
\end{center}
\tablefoot{
The \citet{cenarro2007} super-solar stars are ordered by decreasing metallicity.
The first two columns are the identification of the star, the next three are the atmospheric parameters, and the last is the reference code for these values.
\tablebib{
(1)~\citet{cenarro2007};
(2)~\citet{oinas1977};
(3)~\citet[0.12 dex subtracted on \feh{} accounting for a systematic bias pointed in the original publication]{brown1989};
(4)~\citet{fernandez1990};
(5)~\citet{MW1990};
(6)~\citet[compilation]{W94};
(7)~\citet{luck1995}; 
(8)~\citet{gorgas1999};
(9)~\citet{vanture2002};
(10)~\citet{hekker2007};
(11)~\citet{smith2013};
(12)~\citet{jonsson2014};
(C)~Our compilation;
(PVK)~\citet{prugniel2011};
(M)~Present determination.
}
}
\end{longtab}

\subsection{Super-solar giant stars}

The MILES library contains 12 stars listed in \citet{cenarro2007} as super-solar giants, with \feh$ > 0.1$ and \teff{}~$<~4000$ K. These stars play an important role in the modelling of metal-rich old stellar populations, and we discuss these stars below.
Table~\ref{tab:supersolar} presents the details of our compilation. The first line for each star are the Cenarro parameters, the line with reference "C" is the set of parameters adopted in our compilation, "PVK" are the values used as input to compute the interpolator, and "M" are our final measurements.

For all the 12 stars, the super-solar status results from a single measurement in the literature. Except for one star, our new search leads us to adopt revised parameters that are essentially consistent with a solar (or subsolar) metallicity. This does not come as a surprise, as it is expected that the margins of the parameter space are particularly contaminated by stars with very uncertain or erroneous parameters. In the core of the parameter space, where the compilation is on average reliable, our measurements can detect the outliers. However, at the margins: (i) the interpolator may reflect the bias of the compilation, and (ii) our measurements may be affected by a bias towards the core of the parameter space (although the extrapolation support described in PVK shall essentially prevent this). Therefore, a careful examination of the measurements available for those stars is essential.

Five of these stars were examined in \citet{MW1990}, where the metallicity was measured in two ways. 
\feh{} was measured from high-excitation Fe I lines, which are less sensitive to non-LTE effects that 
are important in red giants, and from both high- and low-excitation lines using 72 Cyg as a standard (and 
assuming the non-LTE effect is similar in the target and in the standard). The first method, in 
principle, should be preferred, but the measurements were based on only a handful of weak lines, while the latter 
method could use a large number of lines. The Pastel database and, following it, \citet{cenarro2007}, adopted the measurements 
from the first method only. We adopted the average of the two measurements, and report the values in 
Table~\ref{tab:mw1990}. In addition, for HD\,120933, \citet{MW1990} assumed a temperature that is warmer than our adopted 
value by 200 K; with our \teff, \feh{} would have been lower. We corrected this bias when adopting our compiled value.

HD\,175865 (M5 III, MILES~705) is a semi-regular pulsating star and the parameters in our compilation for this star
are [$3297, 0.50, -0.25$] where \teff{} is averaged from \citet{mozu2003} (\teff{}\,=\,3174\,K), derived from interferometric angular diameter, and flux from 11-colour photometry,
and \citet{W94} (\teff=$3420$\,K). 
\logg = 0.5\,dex is from the \citet{W94} compilation and was adopted by \citet{gorgas1999} and \citet{cenarro2007}.
\feh{} was computed by \citet{gorgas1999} from the Lick index Fe5270 using the fitting functions of \citet{W94}. Since they used a temperature 400 K warmer than that adopted here, their \feh{} should be decreased by about 0.4 dex.
We did not find any other measurement in the literature.
In PVK, the estimated parameters are [$3181, 0.47, -0.29$] and we estimate [$3316, 0.36, 0.06$] using V2.

\begin{table}
\caption{Metallicities from \citet{MW1990} for the metal-rich cool giants}
\label{tab:mw1990}
\begin{tabular}{lcrrr}
\hline\hline
\multicolumn{1}{c}{Name} & \multicolumn{1}{c}{MILES} & \multicolumn{1}{c}{\feh} & \multicolumn{1}{c}{\feh} & \multicolumn{1}{c}{\feh} \\
     &       & \multicolumn{1}{c}{high}   & \multicolumn{1}{c}{low}    & \multicolumn{1}{c}{mean}  \\
\hline
HD120933 & 487 & 0.50$\pm$0.38 &$ 0.07$& 0.28\\
HD013520 & 080 &$-$0.24$\pm$0.15 &$-0.26$&-0.25\\
HD138481 & 560 & 0.20$\pm$0.35 &$ 0.06$& 0.13\\
HD131918 & 530 & 0.22$\pm$0.24 &$ 0.07$& 0.15\\
HD148513 & 597 & 0.04$\pm$0.15 &$-0.03$& 0.00\\ 
\hline
\end{tabular}
\tablefoot{
The first \feh{} column are from high-excitation Fe I lines, the second for both high and low-excitation measured relative to 72 Cyg and corrected for \feh(72 Cyg) = $-0.08$, and the last column is the average value that we adopted.
}
\end{table}

The mean \citet{cenarro2007} \feh{} for the 12 stars is 0.26 dex, our mean compiled value is 0.03 dex, and the average measured value is $-0.03$ dex. This establishes that the metallicity of these stars is consistent with solar. 
For the warmer stars, $4000~<~$\teff{}$~<~4800$\,K, about half of the stars, indicated as super-solar in \citet{cenarro2007}, are reliably very metal rich.

\subsection{Cluster interlopers}\label{sec:interlopers}

We found two stars in the cluster fields that are not cluster members. 
We describe them hereafter:

\begin{description}

\item[MILES~934:] 
\citet{sandquist2010} concluded from proper motion analysis that Cl*NGC6205 SAV A171 is not a member of NGC\,6205 (M\,13).
To check this suggestion, we fitted the spectrum with free \feh{}, and we obtained [$4266, 1.35, -0.80$],
where the metallicity is discordant with that of the cluster (\feh\,=\,$-1.53$\,dex). This confirms that the star is indeed an interloper.

\item[MILES~967:] 
NGC6838 1078 was diagnosed by \citet{bessell1983} as a probable interloper on the basis of its velocity. From temperature-colour relations, the author reports
\teff\,=\,4260\,K and from a spectroscopic analysis \feh\,=\,0.20\,dex and \logg\,=\,0.90\,dex.
In \citet{cenarro2007}, this star is cross-identified with MILES~968
and is located about 10 arcsec apart from NGC6838~1077, cross-identified
with MILES~967. High-resolution spectroscopic studies of NGC6838~1077 returned
consistent atmospheric parameters: 
[$3900, 0.55, -0.80$] \citep{melendez2009}, 
[$4033, 1.09, -0.84$] \citet{carr2009}, 
and  [$4248, 1.25, -0.67$] \citep[from medium resolution spectroscopy; probably less precise]{kirby2008}.
For the two spectra, \#967 and \#968, our fits return the solutions 
[$4261, 1.87, -0.22$] and [$3952, 1.46, -0.81$], respectively. The second one matches the previous measurements for NGC6838~1077 and we believe that the two targets were swapped.
After correcting the identifications of \#967 and \#968, we confirm the earlier suspicion that NGC6838~1078 = \#967 is not a member of the cluster.

\end{description}

\begin{table*}
\begin{center}
\caption{Derived atmospheric parameters of 46 cluster stars from our sample.}
\begin{tabular}{lcrllr}
\hline\hline
Name & MILES & Cluster  & \multicolumn{3}{c}{Derived parameters}  \\
     & No.$^a$ & \feh{}$^b$  &  \teff{}  & \logg{}  &    \feh{}      \\
\hline
NGC288 77                  & 897  & $-1.32$ &   4224 &    1.08 &  $   -1.35 \pm  0.05 $  \\\hline
NGC1904 153$^c$            & 915  & $-1.60$ &   4251 &    0.74 &  $   -1.64 \pm  0.05 $  \\
NGC1904 160$^c$            & 916  & $     $ &   4212 &    0.59 &  $   -1.73 \pm  0.04 $  \\
NGC1904 223                & 917  & $     $ &   4141 &    0.64 &  $   -1.69 \pm  0.05 $  \\\hline
NGC2420 140                & 918  & $-0.31$ &   4440 &    1.95 &  $   -0.28 \pm  0.03 $  \\\hline
NGC2682 108                & 919  & $ 0.00$ &   4202 &    2.12 &  $   -0.07 \pm  0.03 $  \\\hline
Cl*NGC5272 S I-IV-25       & 923  & $-1.50$ &   4453 &    1.03 &  $   -1.48 \pm  0.06 $  \\
Cl*NGC5272 S I-III-28      & 924  & $     $ &   4261 &    0.67 &  $   -1.59 \pm  0.04 $  \\
NGC5272 398                & 925  & $     $ &   4575 &    1.14 &  $   -1.51 \pm  0.08 $  \\\hline
Cl*NGC5904 ARP III-03      & 926  & $-1.29$ &   4136 &    0.47 &  $   -1.42 \pm  0.03 $  \\
Cl*NGC5904 ARP IV-19       & 930  & $     $ &   4243 &    0.89 &  $   -1.31 \pm  0.05 $  \\\hline
Cl*NGC6205 SAV B786        & 935  & $-1.53$ &   4140 &    0.46 &  $   -1.42 \pm  0.03 $  \\\hline
NGC6341 4114               & 936  & $-2.31$ &   4584 &    1.16 &  $   -2.53 \pm  0.07 $  \\
NGC6341 3013               & 937  & $     $ &   4231 &    0.39 &  $   -2.29 \pm  0.02 $  \\\hline
HD170820 (Cl*IC 4725 A 141)& 939  & $ 0.03$ &   4475 &    1.37 &  $   -0.02 \pm  0.03 $  \\\hline
Cl*NGC6791 GVZH R4         & 940  & $ 0.42$ &   3384 &    0.19 &  $    0.45 \pm  0.03 $  \\
Cl*NGC6791 GVZH R16        & 942  & $     $ &   3882 &    1.67 &  $    0.26 \pm  0.03 $  \\
Cl*NGC6791 GVZH R19        & 943  & $     $ &   3896 &    1.83 &  $    0.25 \pm  0.03 $  \\\hline
Cl*NGC6838 AH A9           & 944  & $-0.78$ &   4378 &    1.39 &  $   -0.87 \pm  0.06 $  \\
NGC6838 1109               & 945  & $     $ &   4764 &    2.35 &  $   -0.77 \pm  0.13 $  \\
NGC6838 1095               & 946  & $     $ &   4475 &    1.43 &  $   -0.93 \pm  0.05 $  \\
NGC6838 1075               & 948  & $     $ &   4771 &    2.44 &  $   -0.83 \pm  0.13 $  \\
NGC6838 1073               & 949  & $     $ &   4636 &    1.84 &  $   -1.04 \pm  0.18 $  \\
NGC6838 1071               & 951  & $     $ &   4350 &    1.69 &  $   -0.88 \pm  0.08 $  \\
NGC6838 1066               & 954  & $     $ &   4210 &    1.55 &  $   -0.88 \pm  0.05 $  \\
NGC6838 1065               & 955  & $     $ &   4648 &    1.94 &  $   -0.81 \pm  0.10 $  \\
NGC6838 1064               & 956  & $     $ &   4455 &    1.50 &  $   -0.89 \pm  0.08 $  \\
NGC6838 1063               & 957  & $     $ &   4629 &    1.46 &  $   -0.91 \pm  0.09 $  \\
NGC6838 1021               & 958  & $     $ &   4447 &    1.32 &  $   -0.85 \pm  0.06 $  \\
NGC6838 1037               & 959  & $     $ &   4498 &    1.96 &  $   -0.88 \pm  0.09 $  \\
NGC6838 1009               & 961  & $     $ &   4779 &    1.97 &  $   -0.76 \pm  0.12 $  \\
NGC6838 1053               & 964  & $     $ &   4609 &    1.44 &  $   -0.92 \pm  0.09 $  \\
NGC6838 1077               & 968  & $     $ &   3952 &    1.46 &  $   -0.81 \pm  0.04 $  \\
Cl*NGC6838 AH S            & 969  & $     $ &   4239 &    1.27 &  $   -0.79 \pm  0.04 $  \\
Cl*NGC6838 AH I            & 971  & $     $ &   4212 &    1.41 &  $   -0.80 \pm  0.04 $  \\\hline
NGC7789 329                & 972  & $ 0.01$ &   5030 &    2.37 &  $   -0.07 \pm  0.04 $  \\
NGC7789 468                & 973  & $     $ &   4156 &    1.71 &  $   -0.02 \pm  0.02 $  \\
NGC7789 353                & 975  & $     $ &   4520 &    2.25 &  $   -0.02 \pm  0.04 $  \\
NGC7789 415                & 976  & $     $ &   3815 &    1.17 &  $    0.04 \pm  0.02 $  \\
NGC7789 461                & 977  & $     $ &   4110 &    1.68 &  $   -0.04 \pm  0.02 $  \\
NGC7789 501                & 978  & $     $ &   4049 &    1.64 &  $   -0.03 \pm  0.02 $  \\
NGC7789 575                & 979  & $     $ &   4509 &    2.10 &  $   -0.05 \pm  0.03 $  \\
NGC7789 637                & 980  & $     $ &   4815 &    2.45 &  $   -0.04 \pm  0.05 $  \\
NGC7789 765                & 981  & $     $ &   4348 &    1.93 &  $   -0.08 \pm  0.03 $  \\
NGC7789 859                & 982  & $     $ &   4548 &    2.24 &  $   -0.15 \pm  0.04 $  \\
NGC7789 971                & 985  & $     $ &   3746 &    1.22 &  $    0.00 \pm  0.02 $  \\\hline

\hline
\end{tabular}
\label{tab:cluster_stars}

\tablefoot{\tablefoottext{a}{Identification number in \citet{cenarro2007}.}
\tablefoottext{b}{Cluster \feh{} from Tables~\ref{tab:compil_globularclusters} and~\ref{tab:compil_openclusters}.\\
\tablefoottext{c}{Used only blue part of the spectrum in the range [3600,5500] \AA. Beyond 5500 \AA{}, spectrum is not well fitted.}\\}
}
\end{center}
\end{table*}

\subsection{Cluster stars}
\label{sec:clusters}

The MILES library contains 89 stars in the field of 17 different clusters that were determined to be important to achieve an extended coverage of parameter space. 
Fifty of these stars are in our selection (the others are warmer),
but two of them were found to be non-members (see Sect.~\ref{sec:interlopers}). In the previous sections, \teff{} and \logg{} of these stars were measured, assuming their metallicity is the mean metallicity of the cluster.
In this section, we reanalyse these 46 stars (ignoring MILES 941 and 927 discussed in Sect.~\ref{sec:outliers})
, leaving three parameters free. This allows us to further estimate the precision of our metallicity determinations.

The determined parameters are reported in Table~\ref{tab:cluster_stars}.
Table~\ref{tab:measured_cluster_feh} gives the mean metallicity obtained for each cluster after averaging the individual measurements. In general, these values agree with the compilation presented in Tables~\ref{tab:compil_globularclusters}~and~\ref{tab:compil_openclusters}.

\begin{figure}
\centerline{\includegraphics[scale=0.6]{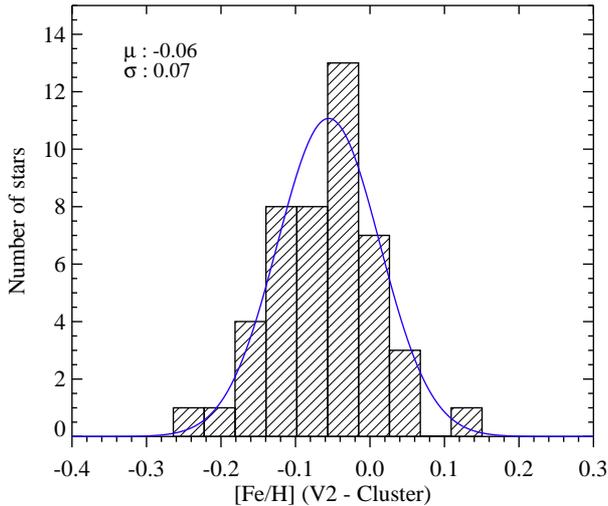}}
\caption{Histogram of metallicity residuals for the cluster stars.
The X-axis is the difference between the measured and compiled \feh, and Y-axis shows the number of stars.
The blue line is the Gaussian with the same mean and standard deviation as the distribution.
}
\label{fig:hist_ngc}
\end{figure}

\begin{table}
\caption{
Mean measured metallicities of the clusters.}
\begin{tabular}{lcrcr}
\hline\hline
Cluster & N & \multicolumn{1}{c}{\feh{}} & \multicolumn{1}{c}{$\sigma(\feh{})$} & \multicolumn{1}{c}{$\Delta(\feh{})$} \\
\hline
NGC288    & 1    & $  -1.35 \pm    0.05$ &    -   & $  -0.03$ \\
NGC1904   & 3    & $  -1.69 \pm    0.05$ &   0.04 & $  -0.09$ \\
NGC2420   & 1    & $  -0.28 \pm    0.03$ &    -   & $   0.03$ \\
NGC2682   & 1    & $  -0.07 \pm    0.03$ &    -   & $  -0.07$ \\
NGC5272   & 3    & $  -1.53 \pm    0.06$ &   0.05 & $  -0.03$ \\
NGC5904   & 2    & $  -1.37 \pm    0.04$ &   0.06 & $  -0.07$ \\
NGC6205   & 1    & $  -1.42 \pm    0.03$ &    -   & $   0.11$ \\
NGC6341   & 2    & $  -2.41 \pm    0.05$ &   0.12 & $  -0.10$ \\
IC 4725   & 1    & $  -0.02 \pm    0.03$ &    -   & $  -0.05$ \\
NGC6791   & 3    & $   0.32 \pm    0.03$ &   0.09 & $  -0.10$ \\
NGC6838   & 17   & $  -0.86 \pm    0.08$ &   0.07 & $  -0.08$ \\
NGC7789   & 11   & $  -0.04 \pm    0.03$ &   0.05 & $  -0.05$ \\

\hline
\end{tabular}
\label{tab:measured_cluster_feh}

\tablefoot{
The column labelled N gives the number of stars measured in the cluster; 
\feh{} is the mean of the measured metallicity, and the error is the mean
measured error;
$\sigma(\feh{})$ is the standard deviation of the measurements (for the clusters
where more than one star were measured);
$\Delta(\feh{})$ is the bias computed as the mean measured \feh{} minus
the compiled value.\\
}
\end{table}

The histogram of the deviations from the compiled \feh{} is presented in Fig.~\ref{fig:hist_ngc}. A Gaussian of the same mean and standard deviation is over-plotted on the histogrammed data. 
The mean residual and dispersion for the 46 stars are $-0.06$\,dex and $0.07$\,dex, respectively, and the mean estimated error on \feh{} is $0.06$ dex. This confirms that the estimated error on \feh{} has the right magnitude.

   \section{Conclusions}\label{sec:conclusions}
   
We analysed the spectra of 331 MILES stars cooler than approximately 4800~K (K- and M-type stars) to improve the interpolator previously presented in PVK and to refine the determination of atmospheric parameters of these stars.

The new interpolator (V2) extends the validity range of the previous version towards the M-type stars, and the biases between the measurements of \teff, \logg, and \feh{} and the reference values compiled from the literature tend to be reduced.
We therefore conclude that the new interpolator is a valuable improvement, and we deliver a new homogeneous set of atmospheric parameters for the cool MILES stars.

\begin{table*}
\begin{center}
\caption{Summary of important notices about some MILES spectra.}
\label{tab:remarks}
\begin{tabular}{clll}
\hline\hline
MILES & Original name & Revised name & Remark \\

\hline
175 & MS 0515.4-0710 & & Cross-identified with V1855 Ori\\
501 & HD124897 & & (Arcturus) The blue segment corresponds to a wrong source \\
591 & Gl627B   & HD147379  & Certainly rather HD147379 = Gl617A\\
593 & HD147923 & HD147923? & Probably a pointing error\\
720 & HD187218 & & The spectrum could no be fitted; excluded from the sample \\
927 & Cl*NGC5904 ARP II-51 & & Low quality; suspected contamination of pointing error\\
934 & Cl*NGC6205 SAV A171 & &non-member\\
952 & Cl*NGC6838 KC 263& & The blue segment corresponds to a wrong source; excluded \\
964 & NGC6838 1053 & &Contaminated by NGC6838 1052\\
967 & NGC6838 1077 & NGC6838 1078 & Inverted with \#968; non-member\\
968 & NGC6838 1078 & NGC6838 1077 & Inverted with \#967 \\
\hline
\end{tabular}
\end{center}
\end{table*}

For the coolest stars of the sample (\teff{}\,$\leq$\,3800\,K), the mean external precision ($1/\sqrt{2}$ times the standard deviation between our measurements and the compiled values from the literature) on the 
\teff{} measurements improves to 63\,K with V2 as compared to 93\,K in PVK. 
We corrected the systematics observed in the temperature of the coolest stars and in the metallicity of the most metal-poor stars. 
For the other \teff{} ranges, there is no significant change in the precision, but we improved the systematics as compared to PVK (see Table~\ref{tab:biases}). 
We also identified and corrected anomalies for eleven stars, including wrong or imprecise identifications, corrupted spectra or cluster interlopers.

The temperature determinations are reliable down to \teff\,$\approx$\,2900 K (M6 for the dwarfs and M7 for the giants).
Above about 4000~K, the interpolator is robust over an extended range of metallicities, down to \feh\,$\approx\,-2.5$ dex for the giants and to \feh\,$\approx\,-0.7$ dex for the dwarfs. For the giants, the metallicities are reliable up to \feh $\approx 0.40$ dex.
In the range $3300\,<\,$ \teff\,$<\,4000$~K, the metallicity coverage narrows to  \feh\,$>\,-0.6$~dex. For the giants, the metallicity measurements have a low reliability for \teff\,$<\,3800$~K.

The limitations in the validity range of the interpolator are essentially determined by the intrinsic coverage of the parameter space by the library.
For the coolest giants, the lack of \feh{} measurements for some stars included in the library is an additional limitation.

Although the new interpolator has been extensively studied for the cool stars only, we tested that its quality is comparable to the previous version over the rest of the range of parameters. This was made by splitting the parameter space into small regions, as in Table~\ref{tab:biases}. The biases were not amplified anywhere, and the comparison of the measurements performed with both versions shows they are compatible within the error bars.
This interpolator shall, therefore, be used as a replacement for the version delivered in PVK.

The parameters listed in Table~\ref{tab:params} are restricted to the current subsample of MILES, and we have shown that they are generally more reliable than those presented in PVK. 
In order to assess whether the present series can be combined with PVK to supply the parameters for the whole sample, we compared the measurements and the error estimates of the two series for the 100 stars in the range 4500\,<\,\teff{}\,<\,4800~K.
We observe an average difference (PVK-V2)
of $\sim$ 30\,K on \teff{}, $0.05$\,dex on \logg{} and $0.04$\,dex on \feh{} between the PVK and V2 series. The corresponding estimates of standard deviation are 42\,K, $0.08$\,dex and $0.04$\,dex, respectively, and the external errors estimated in PVK and V2 are comparable. The small discontinuity between the two sets is significant, in particular for \teff, and should be corrected if one needs to join the two series.

In the recent years, the quality of the spectra and the characterization of the cool stars has progressed significantly, and future spectral libraries should sustain further improvements by including a larger number of cool stars. These developments are needed for stellar population studies, as well as for the study and characterization of cool stars.

The new interpolator is publicly available, and can be used with the ULySS package. The file can be downloaded either from Vizier or from the ULySS web site.

\begin{acknowledgements}

KS acknowledges the University Teaching Assistantship provided by the University of Delhi, Delhi, India and Centre de Recherche Astrophysique de Lyon (CRAL), and Observatoire de Lyon for supporting a visit to Lyon where this work was completed. HPS thanks Centre de Recherche Astrophysique de Lyon (CRAL), Observatoire de Lyon, Universit\'{e} Claude Bernard for Invited Professorship, and a research grant from University of Delhi. 
The authors also thank the anonymous referee for valuable comments/suggestions, which led to significant 
improvement of the work presented in this paper.
This research has made use of the SIMBAD data base (operated at CDS, Strasbourg, France) and the NASA's Astrophysics Data System Article Service.
This research also used data from the ESO Science Archive Facility (FEROS, HARPS and X-Shooter, \url{http://archive.eso.org}), from the Sloan Digital Sky Survey (SDSS; \url{http://www.sdss.org}), and from the Observatoire de Haute-Provence archive (\url{http://atlas.obs-hp.fr}).

\end{acknowledgements}

   \bibliographystyle{aa} 
   \bibliography{aa}  

\begin{appendix} 

\section{Characteristics of the current version of MILES}\label{appendix:miles91}

\begin{figure*}
\begin{center}
\includegraphics{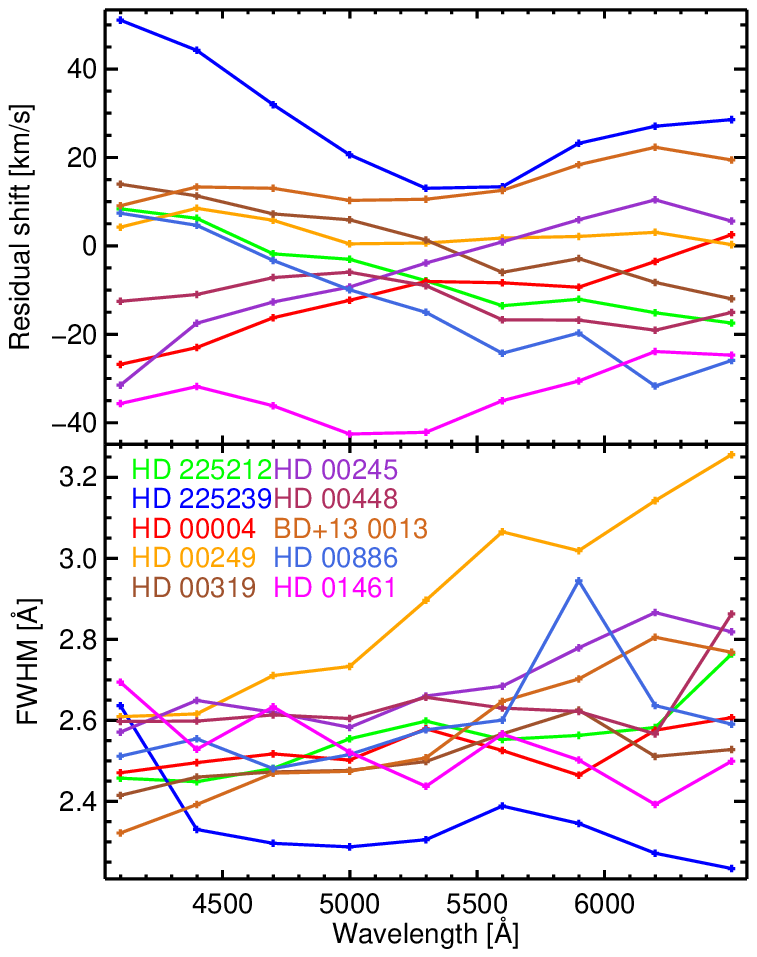}
\includegraphics{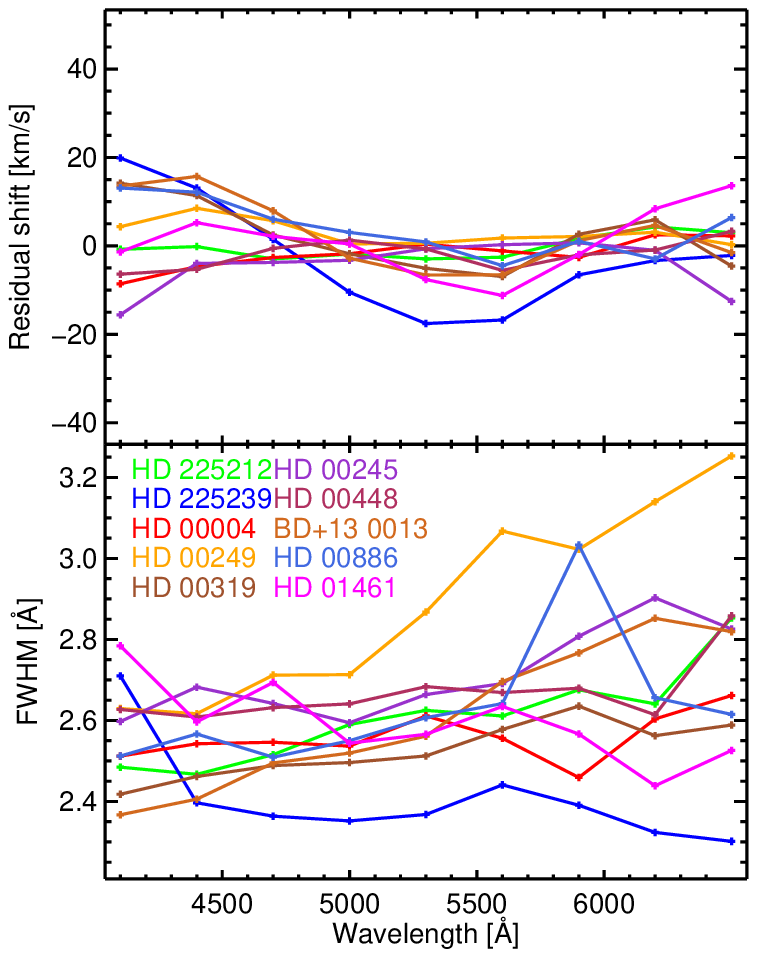}
\caption{Line spread function of some MILES spectra. The first ten stars of the library are represented with different colours, as labelled. The left panels are for MILES v9.0 (the first release) and the right panels for MILES v9.1 (the updated version). The top panels show the velocity shift, i.e. wavelength calibration  error, as a function of the wavelength, and the bottom panels present the FWHM resolution.}
\label{fig:indiv_lsf}
\end{center}
\end{figure*}

Other than measuring the atmospheric parameters of the MILES stars, PVK also made an assessment of the wavelength calibration and line-spread function (LSF) of this library. It appeared that the wavelength calibration of some stars was imprecise, and in other cases, the error on the velocity rest-frame reduction exceeded one pixel. These errors were corrected within the procedure that computed the interpolator.
Soon after the publication of PVK, the authors of MILES released a new version of the library, in which they corrected the wavelength issue \citep{falconbarroso2011}. 
The initial release is called version 9.0, and the updated is version 9.1.
In this appendix, we briefly assess the characteristics of the new version.

We repeated the tests carried out in PVK. The LSF describes
how infinitely narrow spectral lines are transformed by the instrument and data reduction process. We modelled it with a Gaussian whose shift expresses the wavelength calibration error, and whose dispersion measures the spectral resolution. As both the wavelength calibration error and the resolution can change with the wavelength, we measured the LSF in a succession of segments along the wavelength axis. The measurements were done by comparing each MILES spectrum with various templates, including ELODIE spectra and synthetic spectra from the \citet{coelho2005} library.

In Fig.~\ref{fig:indiv_lsf}, we
reproduce Fig. 1 of PVK to compare the two versions of MILES. 
It presents the LSF of the first ten stars of the library. In the original version, two of the represented stars had systematic shifts of about 40 \kms, probably due to imprecise reduction of the spectra to the velocity rest frame (the correction was performed using catalogued proper velocities). 
The rest of these ten spectra displayed drifts of the shift by over 30~\kms,
reflecting imprecise wavelength calibrations resulting from flexure or thermic effects in the spectrograph.
With the new version (right panel) the calibration appears greatly improved.
On average, over the library, the wavelength calibration errors accounted for a dispersion of 14 \kms{} in the initial release, and with the new version it is reduced to 8~\kms.

\begin{figure}
\includegraphics{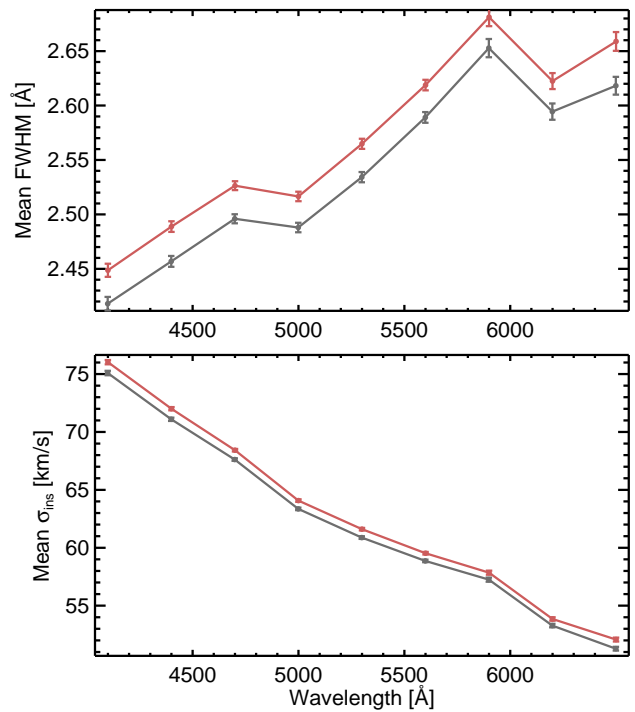}
\caption{Mean resolution of the MILES library. The red line and symbols present the mean resolution of the updated version of the library (v9.1) as a function of the wavelength, and the grey line is for the original version (v9.0). The top panel shows the FWHM resolution, and the bottom panel indicates the instrumental velocity dispersion, $\sigma_{ins}$ (FWHM = $2.355 \lambda \sigma_{ins}/c$, where $\lambda$ is the wavelength and $c$ is the speed of light).}
\label{fig:mean_lsf}
\end{figure}

Figure.~\ref{fig:mean_lsf} shows the resulting mean resolution as a function of the wavelength. With the new version, the resolution at 5300~\AA{} is 2.56~\AA{} FWHM.
It is slightly degraded with respect to the original version where it was 2.53~\AA{} FWHM. The degradation results from the fact that the correction was implemented as an additional rebinning.

Altogether, the new MILES version appears more homogeneous and precise in terms of wavelength calibration, but its resolution is slightly degraded. When spectra are combined together to make an interpolator or assemble stellar populations, the two effects compensate each other and the two versions are almost equivalent.

We chose to use the latest MILES v9.1 because the homogeneity of the wavelength
calibration is more important for the measurements of the atmospheric parameters than the resolution. 
Indeed, \cite{koleva2012} showed that lowering the resolution has only a small effect on the determinations. By contrast, a drift of 30 \kms{} over the whole wavelength range results in a systematic underestimate of the metallicity by 0.02 dex (for a 4500\,K star); this is smaller than other systematics shown in the present paper, but it is comparable to the errors due to noise. 

\subsection{Number of independent pixels in MILES spectra}

Other than affecting the spectral resolution, the rebinning introduces a correlation
between adjacent pixels. Consequently, the number of independent 
pixels is reduced, which affects the error estimates (the fitting
errors are inversely proportional to the square root of the number of degrees of freedom). 
Although no error spectra are given in MILES, the assumption that
the fitting residuals are entirely due to a Gaussian noise gives an upper 
limit to the errors due to noise (see Sect.~\ref{sec:error_analysis}), providing that the number of degrees of freedom is known. We therefore assess this below.

Although the technical details of the data processing of the MILES library are not precisely known, it is clear that the data had to be rebinned once to obtain a two-dimensional frame whose axes were the wavelength and the position along the slit. This frame was used to subtract the background and extract the one-dimensional spectrum. Later on, the spectrum was again rebinned into the rest frame, and a further rebinning was applied to produce the second release.

A rebinning can 
be decomposed mathematically in three operations: (i) integrating the spectrum 
as a function of the wavelength, (ii) interpolating this integral at the edges
of the new bins, and (ii) differentiating to obtain the flux in each new bin\footnote{At variance, a resampling interpolates the signal at new points. In some circumstances, a resampling is an acceptable approximation for rebinning}.
By definition, this operation preserves the total flux (but not the flux 
distribution). The choice of the interpolation schema determines the effect on the resolution and correlation.
In the present case, the rebinnings where done preserving the mean spectral dispersion of the 
original data (0.9 \AA/pix). It means that each time the flux of a pixel
was redistributed on one, two, or at most three pixels (the latter case when the
rebinned pixel is smaller than the original one).

The effect on the correlation between adjacent pixels depends on the detailed 
configuration. When one pixel is projected to exactly one pixel, no new correlation is introduced, but when one pixel is equally shared between two new ones (e.g. a half-pixel shift), a correlation length of two arises. The latter is the strongest effect. Since the rebinning operations made in MILES are not mere global shifts of the grid, the correlation effect varies along the spectrum from no correlation to a two-pixels correlation. We cannot reproduce the details, but on average, we can assume that each rebinning is equivalent to a shift of 1/4 (the effect is symmetrical around 1/2 pixels shift). The actual effect on the resolution and correlation is determined by the interpolation schema that was chosen.

\begin{figure}
\includegraphics{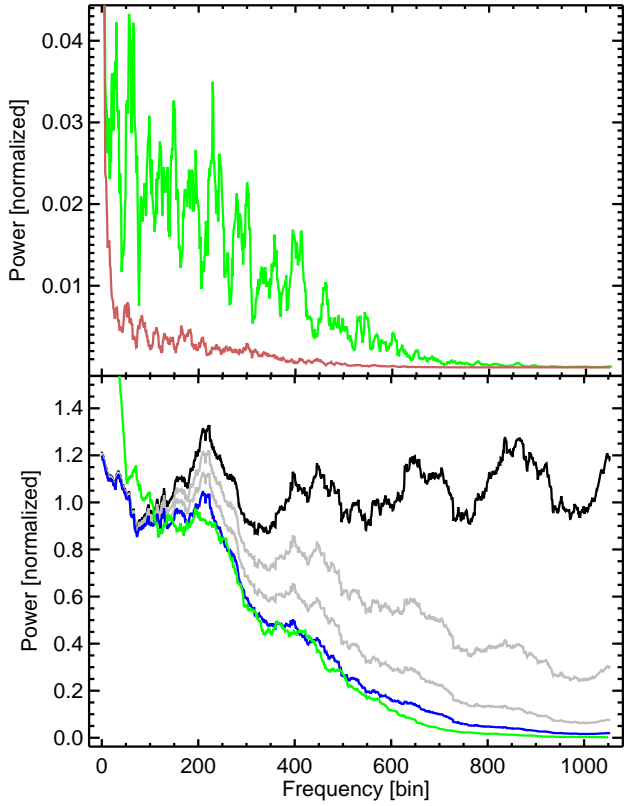}
\caption{Effect of rebinning on the power spectrum. 
(a) The top panel shows the power spectra of MILES 915 (a noisy spectrum, in green) and MILES 924 (a high S/N spectrum, red); the power spectra are smoothed with a 10 bin boxcar for visualization purposes.
(b) The bottom panel represents the power spectrum of white noise (in black), and of its transformation after one, two (in grey) and three (in blue) quarter-pixel shift linear rebinnings. The green line is the difference between the two spectra of the top panel. The power spectra are smoothed with a 100 bin boxcar for visualization purposes.}
\label{fig:power}
\end{figure}

In order to determine the magnitude of the correlation introduced by the data processing, we are considering two MILES spectra. MILES 915, is the lowest
S/N observation of the present sample, and MILES 924 is a similar star, but with 
high S/N. Figure~\ref{fig:power} shows the power spectrum of the red segment of these two observations. MILES 915 is represented in green, and MILES 924 in red. The damping of the high frequencies in both spectra is primarily due to the resolution of the spectrograph, but the power excess of MILES 915 is due to the difference of noise level between the two observations.

The bottom panel of Figure~\ref{fig:power} shows the power spectrum of the difference between the spectra of MILES 915 and 924 (in green), which is therefore essentially the noise. It is superposed to a simulated white noise (in black) and to this noise after one, two, and three consecutive quarter-pixel rebinning. The latter one, in blue, tightly matches the noise. Each rebinning decreases the number of independent pixels, accumulating a reduction factor of 1.60, 2.16, and 2.69 after each step. 

We therefore conclude that the data processing of MILES reduced the number of independent pixels by a factor of about 2.6.
For our analysis, we assume that the number of independent pixels is reduced by the factor determined above. In the data model of ULySS, this information is coded as the {\tt dof\_factor} meta-data of a spectrum.

\end{appendix}
   
\label{lastpage}
\end{document}